\documentclass[11pt,prd,superscriptaddress,nofootinbib]{revtex4}
\usepackage[english]{babel}
\usepackage{graphicx}
\usepackage{mathtext}
\usepackage{indentfirst}
\usepackage{epsfig,amsmath,amsfonts}
\usepackage{amsmath,amssymb,braket}

\newcommand{\beq}[1]{\begin{eqnarray}\label{#1}}
\newcommand\eeq {\end{eqnarray}}
\newcommand\bqa {\begin{eqnarray}}
\newcommand\eqa {\end{eqnarray}}

\newcommand{\eq}[1]{(\ref{#1})}
\newcommand{\bear}{\begin{array}}
\newcommand{\enar}{\end{array}}

\newcommand{\A}{{\cal{A}}}

\begin{document}
\begin{titlepage}
\hfill ITEP-TH-13/15\hspace*{3.4mm}

\hfill DAMTP-2015-46\hspace*{0.7mm}
\vspace{2.5cm}
\begin{center}

\centerline{\Large \bf Hawking radiation and secularly growing loop corrections}

\vskip 1.5cm {Emil T.\ Akhmedov$^{1,2}$, Hadi Godazgar$^{3}$ and
Fedor K.\ Popov$^{1,2,4}$}
\\
{\vskip 0.5cm
$^{1}$International Laboratory of Representation Theory and Mathematical Physics,\\
National Research University Higher School of Economics,\\
Russian Federation
\vskip 0.5cm
$^{2}$B. Cheremushkinskaya, 25,
Institute for Theoretical and Experimental Physics,\\
117218, Moscow, Russian Federation}
\vskip 0.5cm
$^{3}$DAMTP, Centre for Mathematical Sciences,
University of Cambridge,\\
Wilberforce Road, Cambridge, CB3 0WA, United Kingdom
\vskip 0.5cm
$^{4}$Institutskii per, 9, Moscow Institute of Physics and Technology,\\
141700, Dolgoprudny, Russian Federation

\vskip 0.5cm
\end{center}

\vskip 0.35cm

\begin{center}
August 27, 2015
\end{center}

\noindent

\vskip 1.2cm

\centerline{\bf Abstract}

\noindent
We study the expectation value of the energy momentum tensor during thin shell collapse for a massive, real, scalar field theory. At tree-level, we find thermal, Hawking-type, behavior for the energy flux. Using the Schwinger-Keldysh technique, we calculate two-loop corrections to the tree-level correlation functions and show that they exhibit secular growth, suggesting the breakdown of the perturbation theory.

\end{titlepage}

\section{Introduction} \label{intro}

Black holes are one of the few laboratories for studying quantum gravity. Indeed, a commonly viewed challenge for any candidate theory of quantum gravity is to provide an understanding of the quantum nature of black holes, and in particular their radiation.

Hawking's semi-classical calculation \cite{Hawking:1974sw} shows that black holes radiate with a thermal spectrum. This has lead to black holes being viewed as objects to which thermodynamic quantities can be associated with, such as temperature and entropy, which satisfy relations analogous to the standard laws of thermodynamics.~\footnote{While these laws were in fact suggested before Hawking radiation was proposed (see reference \cite{Hawking:1973uf} and references therein), they can only be made sense of if black holes radiate.} The thermodynamic perspective on black holes is incredibly attractive and has inspired many questions regarding, for example, a microscopic understanding of its entropy, or whether it implies a holographic interpretation of gravity \cite{tHooft}. However, the thermal spectrum of black hole radiation also poses a puzzle in the form of the information paradox \cite{Hawking:1976ra}.
Our goal, in this paper, is to revisit black hole radiation and study quantum loop corrections to Hawking radiation that are often dismissed as negligible.

Hawking radiation \cite{Hawking:1974sw} (see also \cite{Unruh:1976db}) is a quantum effect that can be seen in a gaussian (non-self-interacting) quantum field theory on a collapse background.~\footnote{For reviews see, for example, references \cite{BD}, \cite{Wipf:1998ss} and \cite{Mottola:2010gp}.} In this paper, we begin by reproducing Hawking radiation in a slightly different setting.

The background that we  would eventually like to consider is a star that is static due to an internal pressure until some moment of time, after which the pressure is switched off and collapse begins. To our knowledge the best model of such a situation is pressure-free collapse of a spherical perfect fluid---Oppenheimer-Snyder collapse \cite{Oppenheimer:1939ue}. For simplicity reasons, however, instead of a ball of the perfect fluid, we study Hawking radiation in the case of a massive thin shell collapse. We will study particle creation during Oppenheimer-Snyder collapse elsewhere.

\begin{figure}
\begin{center}
\includegraphics[scale=0.3]{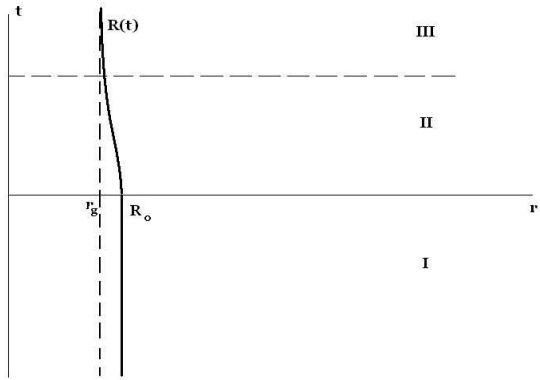}\caption{The collapse of a spherically symmetric thin shell as is seen by an outside Schwarzschild observer. We divide the collapse into three phases: I, II and III}\label{fig1}
\end{center}
\end{figure}

The background in question is described in detail in the section \ref{sec:backgeo}. It has three phases, see figure \ref{fig1}. The phase labeled I in the figure is when the shell is kept fixed at some radius  $R(t) = R_0$ by an additional force. Phase II is a  highly tuned stage of collapse if one wishes to respect spherical symmetry. In fact, if $R_0$ is large enough, because of tidal forces any perturbation violating the spherical symmetry will grow in this phase. Phase III, however, describes the final stage of the collapse, where, without rotation, spherical symmetry is restored (if it was lost in the second stage) because all multipole momenta are radiated away \cite{Price:1971fb}, \cite{Price:1972pw}. This feature of the last stage of collapse is because of the peculiar properties of the horizon---the surface of infinite red shift. This phenomenon is also at the core of the ``No Hair Theorem''. Hence, this stage is universal and does not depend on what is happening during phase II.

It should not be surprising that a non-stationary gravitational background field creates particles. The production rate is then generically non-stationary and non-universal. But at the final stage of the collapse we have a stationary and universal free fall of the shell. Hence, it is natural to expect a stationary and universal particle production rate. Then the question reduces to the following one: What is the spectrum of the particles produced  at the final stage of collapse? In other words, we expect that the free Hamiltonian of a theory on such a time-dependent background should also depend on time. But it is obvious that the free Hamiltonian can be diagonalized before the start of collapse. Its ground state defines the initial conditions for the problem that we study here. Moreover, we expect that it is also possible to diagonalize the free Hamiltonian at the final stage of collapse due to its stationarity. But one can guess that the harmonic functions which do such a diagonalization before collapse do not coincide with those performing the diagonalization at the final stage of collapse. The technical details supporting this general discussion are presented in the section \ref{sec:har}.

In the section \ref{sec:hawtree}, we calculate the time evolution of the expectation value of the stress--energy tensor. We find that before collapse has started the energy flux is zero. But at the final stage of the collapse it is given by a thermal stationary energy flux. We perform the calculation beyond the geometric optic approximation in four dimensional space-time for a massive real scalar field and include all spherical harmonics.

The above discussion is true for a gaussian theory. In section \ref{sec:loop}, we study what happens if self-interactions, in particular a $\phi^4$ interaction, is turned on. The common wisdom is that quantum loop corrections should not change the picture described above in a substantial way. This is probably true for UV corrections. However, in section \ref{sec:loop}, we show that perturbative IR corrections grow with time and, if one considers a long enough period of time they can even dominate over the tree-level contribution. This is, in fact, a well known phenomenon in non-stationary condensed matter theory \cite{LL}, \cite{Kamenev}: UV modes essentially do not feel the presence of the background field, but the behavior of IR modes reveals a change in the background state of the quantum field theory under consideration. In the concluding section we explain how these corrections modify the Hawking flux.

As far as we are aware this is the first study of the effect of interactions on black hole radiation. For other work where various other corrections to the standard Hawking radiation picture are considered see, for example, \cite{Vachaspati:2006ki}, \cite{Brustein:2014faa}, \cite{Brustein:2014iha}, \cite{Saini:2015dea}, \cite{Alberghi:2001cm}, \cite{Takahashi:2010th}, \cite{Khavkine:2010hp}, \cite{Kawai:2015uya} and \cite{Kawai:2013mda}.

\subsection{On the origin of secular growth in non--stationary situation in perturbation theory}

In section \ref{sec:loop}, we show that that the perturbative corrections to the Keldysh propagator of scalars on a collapsing shell background grow secularly. In this subsection we explain why this is not surprising and is to be generically expected in non-stationary situations.

Suppose that one would like to find the time evolution of the expectation value of an operator ${\cal O}$:
\bqa\label{ave}
\left\langle {\cal O} \right\rangle_{t_0}(t) \equiv \left\langle \Psi\left| \overline{T} e^{i\,\int_{t_0}^t dt' H(t')}\,{\cal O} \, T e^{-i\,\int_{t_0}^t dt' H(t')}\right| \Psi\right\rangle.
\eqa
The operator ${\cal O}$ could, for example, be the stress-energy tensor.
Here $H(t) = H_0(t) + V(t)$ is the full Hamiltonian of the theory; while $T$ denotes time-ordering and $\overline{T}$ is anti-time-ordering; $t_0$ is an initial time; $\left|\Psi\right\rangle$ is an initial state. The initial value of the operator  $\left\langle {\cal O} \right\rangle(t_0)$ is known. We also assume some kind of covariant UV regularization.

After transforming to the interaction picture, we get \cite{LL}:
\begin{align}\label{step}
\left\langle {\cal O} \right\rangle_{t_0}(t) &= \left\langle \Psi\left| S^{\dagger}(t, t_0)\, {\cal O}_0(t) \, S(t,t_0) \right| \Psi\right\rangle
\nonumber \\[1mm] &= \left\langle \Psi\left| S^{\dagger}(t, t_0) S^{\dagger}(\infty, t)\, S(\infty, t)\, T\left[ {\cal O}_0(t) \, S(t,t_0)\right]\right| \Psi\right\rangle \nonumber \\[1mm]
&= \left\langle \Psi\left| S^{\dagger}(\infty, t_0)\, T\left[ {\cal O}_0(t) \, S(\infty,t_0)\right]\right| \Psi\right\rangle,
\end{align}
where $S(t_2,t_1) = T e^{-i\,\int_{t_1}^{t_2} dt' \, V_0(t')}$; ${\cal O}_0(t)$ and $V_0(t)$ are the ${\cal O}$ and $V$ operators in the interaction picture. The first step in equation (\ref{step}) is an application of the Baker-Hausdorff formula. To perform the step on the second line of (\ref{step}), we insert the unit operator $1 = S^{\dagger}(\infty, t)\, S(\infty, t)$, which allows us to extend the original time-evolution from $t_0$ to $t$, $S(t,t_0)$, and back, $S^{\dagger}(t,t_0)$, to that which goes from $t_0$ to future infinity, $S(\infty, t_0)$, and back, $S^{\dagger}(\infty, t_0)$. Finally, we place the operator ${\cal O}_0(t)$ on the part of the time contour that is going forward.

Now let us slightly change the problem. Suppose we adiabatically turn on an interaction term, $V$, after time $t_0$. Then the state $\left|\Psi \right\rangle$ does not change before $t_0$. In this case, we can rewrite the expectation value (\ref{step}) as follows:
\bqa\label{calo}
\left\langle {\cal O} \right\rangle_{t_0}(t) = \left\langle \Psi\left| S_{t_0}^{\dagger}(\infty, -\infty)\, T\left[ {\cal O}_0(t) \, S_{t_0}(\infty,-\infty)\right]\right| \Psi\right\rangle.
\eqa
An important question is if one can take $t_0$ to past infinity, $t_0 \to -\infty$. If the answer is positive, then the remaining problem is if this can be done within perturbation theory or non--perturbatively. A well-know situation where this can be done within perturbation theory is when the free Hamiltonian, $H_0$, does not depend on time and is bounded from below. Moreover, $\left|\Psi \right\rangle$ should coincide with its true ground state, $\left|vac \right\rangle$, so that $H_0 \, \left|vac \right\rangle = 0$. In the calculation that follows, we also assume that the interaction term is adiabatically switched off at future infinity---after the time $t$.

In fact, if the state $\left|vac \right\rangle$ is the true vacuum state of the free theory, then, by adiabatically turning on and then switching off the interactions, we cannot excite such a state, i.e.\ $\left\langle vac \left| S^{\dagger}(\infty, -\infty)\right| excited \,\, state\right\rangle = 0$, while $\left|\left\langle vac \left| S^{\dagger}(\infty, -\infty)\right| vac \right\rangle\right| = 1$. Thus, the dependence on $t_0$ disappears at every order of perturbation theory and we arrive at expressions that only contain time-ordering: we obtain an expression that can be calculated with the use of the standard Feynman diagrammatic method:
\begin{align}\label{43}
\left\langle {\cal O} \right\rangle_{t_0}(t) &= \sum_{state} \left\langle vac \left| S^{\dagger}(\infty, -\infty)\right| state \right\rangle \, \left\langle state \left| T\left[ {\cal O}_0(t) \, S(\infty,-\infty)\right]\right| vac \right\rangle \nonumber \\ &= \left\langle vac \left| S^{\dagger}(\infty, -\infty)\right| vac \right\rangle \, \left\langle vac \left| T\left[ {\cal O}_0(t) \, S(\infty,-\infty)\right]\right| vac \right\rangle \nonumber \\[2mm]
&=\frac{\left\langle vac \left| T\left[ {\cal O}_0(t) \, S(\infty,-\infty)\right]\right| vac \right\rangle}{\left\langle vac \left| S(\infty, -\infty)\right| vac \right\rangle} .
\end{align}
In the first line in equation (\ref{43}), we have inserted the unit operator $1 = \sum_{state} \left| state \right\rangle \, \left\langle state \right|$, where the sum runs over a complete basis of states. In the second line, we have used the fact that $\left|vac \right\rangle$ is the only state in the sum that gives a non-zero contribution.

If, however, $\left|\Psi\right\rangle$ is not a vacuum state, or $H_0$ is time dependent (and, hence, cannot be diagonalized once and forever), or $H_0$ is not bounded from below, the above technique cannot be used and we have to directly evaluate the expression given on the right handside of equation \eq{calo}. Moreover, in principle, the dependence on $t_0$ can remain and it frequently does remain within perturbation theory and sometimes even
non–perturbatively---after resummation of leading corrections from all loops. The well known situation when the dependence on $t_0$ does disappear after such a resummation is when the system reaches thermal equilibrium
from a non–stationary initial state \cite{LL}, \cite{Kamenev}. This happens in flat space–time without background fields, when $H_0$ is time--independent and is bounded from below. A good question, however, is whether there is a stationary state in the presence of an eternally acting background field---such as for example global de Sitter space, constant electric field or eternal Schwarzschild black hole. Let us explain how the dependence on $t_0$ reveals itself.

The standard method to calculate an expectation value such as that given in equation (\ref{calo}) is the so-called Schwinger-Keldysh diagrammatic technique, where  both $S$ and $S^{\dagger}$ have to be perturbatively expanded under the quantum average \cite{LL}, \cite{Kamenev}. In such a case every vertex in the diagrams carries either a ``$+$'' or ``$-$'' index, depending on whether
it comes from $S$ or $S^{\dagger}$. Furthermore, after Wick contractions every field, e.g.\ $\phi$, is described by a matrix of propagators, whose entries are given by
\bqa \label{Dpmprop}
D_{-+}\left(X_1,X_2\right) = \langle \phi(X_1) \, \phi(X_2) \rangle, \quad  D_{+-}\left(X_1,X_2\right) = \langle \phi(X_2) \, \phi(X_1) \rangle, \nonumber \\ D_{++}\left(X_1,X_2\right) = \langle T \, \phi(X_1) \, \phi(X_2) \rangle, \quad
D_{--}\left(X_1,X_2\right) = \langle \overline{T} \,\phi(X_1) \, \phi(X_2) \rangle.
\eqa
Propagators with time or anti-time-ordering or without orderings appear depending on whether the field $\phi$ comes from $S$ or $S^{\dagger}$. From their definitions, it is clear that these propagators obey the relation $D_{+-} + D_{-+} = D_{++} + D_{--}$.

After Keldysh rotation \cite{LL}, \cite{Kamenev}, one of the entries of the propagator matrix is set to zero. The other entries are given by the retarded, advanced and Keldysh propagators. The retarded propagator is given by
\begin{align}
D^{R}\left(X_1, X_2\right) &= \theta\left(\Delta t_{12}\right)\, \left[D_{-+}\left(X_1, X_2\right) - D_{+-}\left(X_1, X_2\right)\right] \nonumber \\[2mm]
&= \theta\left(\Delta t_{12}\right) \, \left[\phi(X_1), \phi(X_2)\right], \quad \Delta t_{12} = t_1 - t_2.
\end{align}
While the advanced propagator, $D^A$, is conjugate to the retarded one. Since the commutator $[\cdot,\cdot]$ is a c-number, at tree--level these two propagators do not depend on the state---they define the spectrum of excitations in the theory.

The last entry of the propagator matrix is the Keldysh propagator:
\bqa\label{Keldprop}
D^K\left(X_1, X_2\right) = \frac{1}{2} \left[D_{-+}\left(X_1, X_2\right) + D_{+-}\left(X_1, X_2\right)\right] = \frac{1}{2} \, \left\langle \left\{\phi(X_1), \phi(X_2)\right\}\right\rangle,
\eqa
where $\{\cdot, \cdot\}$ is the anti-commutator. The Keldysh propagator is sensitive to the time evolution of the background state. This is an important fact for finding the perturbative secular growth that we are after in this paper.

To explain the origin of the secular growth, consider the simple example of a spatially homogeneous but non-stationary situation.~\footnote{Of course in the case of spherically symmetric collapse background the details will be different, but conceptually the phenomenon is the same. The details for the collapse situation are presented in the section \ref{sec:loop}. Here we describe a simpler situation to give a flavor of the physical meaning of secularly growing quantum corrections. It is also worth stressing that in standard condensed matter situations one considers a non-stationary initial state rather than a background field.} Then, in the case of a real scalar field $\phi$ we have the following harmonic expansion:
$$
\phi(t,\vec{x}) = \int \frac{d^3\vec{k}}{(2\pi)^3} \, \left[a_{\vec{k}} \, g_k(t) \, e^{i \, \vec{k} \, \vec{x}} + c.c.\right],
$$
where the harmonic functions, $g_k(t) \, e^{i \, \vec{k} \, \vec{x}}$, solve the corresponding Klein-Gordon equation in a time-dependent spatially homogeneous background field in flat space--time. It is not hard to see that if in equation (\ref{Keldprop}) we take the average with respect to an arbitrary state respecting spatial homogeneity, then:
\bqa\label{examDK}
\int d^3\vec{x} \, e^{- i \, \vec{k} \, \vec{x}} \, D^K\left(t_1, \vec{x}; t_2, 0\right) = \left[\frac12 + \left\langle a^{\dagger}_{\vec{k}} \, a_{\vec{k}}\right\rangle\right]\, g^*_k(t_1) \, g_k(t_2) + \left\langle a_{\vec{k}} \, a_{-\vec{k}}\right\rangle \, g_k(t_1) \, g_k(t_2) + c.c. ,
\eqa
where we have used the fact that in spatially homogeneous situations
$$\left\langle a^{\dagger}_{\vec{k}} \, a_{\vec{k}'}\right\rangle = \left\langle a^{\dagger}_{\vec{k}} \, a_{\vec{k}}\right\rangle \, \delta\left(\vec{k} - \vec{k}'\right) \qquad {\rm and} \qquad  \left\langle a_{\vec{k}} \, a_{\vec{k}'}\right\rangle = \left\langle a_{\vec{k}} \, a_{-\vec{k}}\right\rangle \, \delta\left(\vec{k} + \vec{k}'\right).$$
If in such an expression we average over the vacuum, then, $\left\langle a^{\dagger}_{\vec{k}} \, a_{\vec{k}}\right\rangle = 0 = \left\langle a_{\vec{k}} \, a_{-\vec{k}}\right\rangle$. Note that all the standard tree--level calculations of particle fluxes (such as e.g. Hawking flux or Schwinger's electric current) are done with the use of such Wightman correlation functions as (\ref{examDK}) with $\left\langle a^{\dagger}_{\vec{k}} \, a_{\vec{k}}\right\rangle = 0 = \left\langle a_{\vec{k}} \, a_{-\vec{k}}\right\rangle$.

Furthermore, at tree-level $\left\langle a^{\dagger}_{\vec{k}} \, a_{\vec{k}}\right\rangle$ and $\left\langle a_{\vec{k}} \, a_{-\vec{k}}\right\rangle$ remain constant, even if the average is done over an arbitrary state, because all the time dependence of the creation and annihilation operators is absorbed into the harmonic functions $g_k(t)$. Note that the appropriate quantity to define as particle density is $\left\langle a^{\dagger}_{\vec{k}} \, a_{\vec{k}}\right\rangle \, g^*_k(t) \, g_k(t)$, rather than $\left\langle a^{\dagger}_{\vec{k}} \, a_{\vec{k}}\right\rangle$---this can be seen, in particular, from the form of the correlation
function in equation (\ref{examDK}). While for plane waves there is no difference, for the exact harmonics in background fields the difference is obvious. The same comment also holds for the anomalous quantum average $\left\langle a_{\vec{k}} \, a_{-\vec{k}}\right\rangle$. However, below, for simplicity, we refer to  $\left\langle a^{\dagger}_{\vec{k}} \, a_{\vec{k}}\right\rangle$ as particle density and to $\left\langle a_{\vec{k}} \, a_{-\vec{k}}\right\rangle$ as the anomalous quantum average. We hope that this does not cause confusion.

If one turns on interactions, the behavior of the Keldysh propagator, (\ref{examDK}), changes drastically. In particular, $\left\langle a^{\dagger}_{\vec{k}} \, a_{\vec{k}}\right\rangle$ and $\left\langle a_{\vec{k}} \, a_{-\vec{k}}\right\rangle$ are generated, even if they were zero to begin with. That of course is to be expected: the presence of a non-zero anomalous quantum average means that, in a strong field background, the ground state of the theory is changing from past to future infinity once one turns on self-interactions. The same happens with the particle number density. In principle, they will depend on $t_1$ and $t_2$ simultaneously. However, since we are only interested in the leading growing terms, in the limit $t \equiv (t_1+t_2)/2 \to \infty$ and $|t_1 - t_2| = const$, we can neglect the difference between $t_1$ and $t_2$. In fact, in $\lambda \, \phi^4$ theory, the two-loop sunset diagram correction to the Keldysh propagator is given by (\ref{examDK}) with~\footnote{Note that the bubble diagram correction to the propagator, $\sim \lambda$, does not introduce corrections that grow with time because the corresponding contribution is local, i.e. is not sensitive to the position of the external legs.}
\begin{align}\label{aaaa}
\left\langle a^{\dagger}_{\vec{k}} \, a_{\vec{k}}\right\rangle(t) &\propto \lambda^2 \, \iint_{t_0}^t dt_3 \, dt_4 \, \delta\left(\vec{k} + \vec{q}_1 + \vec{q}_2 + \vec{q}_3\right)  g_k\left(t_3\right) \, g_k^*\left(t_4\right) \, \prod\limits_{j=1}^3 \int d^{3}\vec{q}_j \, g_{q_j}(t_3)\, g_{q_j}^*(t_4) , \nonumber \\
\left\langle a_{\vec{k}} \, a_{-\vec{k}}\right\rangle(t) &\propto - \lambda^2 \, \int_{t_0}^t dt_3 \, \int_{t_0}^{t_3} dt_4 \,  \delta\left(\vec{k} + \vec{q}_1 + \vec{q}_2 + \vec{q}_3\right)  g_k^*\left(t_3\right) \, g_k^*\left(t_4\right) \, \prod\limits_{j=1}^3 \int d^{3}\vec{q}_j \, g^*_{q_j}(t_3)\, g_{q_j}(t_4),
\end{align}
if the initial state is chosen to be the ground state of the free Hamiltonian at past infinity:
$a_{\vec{k}} \, \left|ground\right\rangle = 0$. Here $t_0$ is the moment after which $\lambda$ is adiabatically turned on. In spatially inhomogeneous situations the formulae are quite different from the present one, but conceptually the phenomenon is the same. Also if the initial state is not a ground state then (\ref{aaaa}) gets modified.

These formulae are also valid in the stationary situation, when the free Hamiltonian is time independent and is bounded from below. But it is not hard to see that if $g_p(t) \propto e^{- i \, \omega(p) \, t}$, then after change of integration variables from $t_3$ and $t_4$ to $T = (t_3 + t_4)/2$ and $\tau = t_3 - t_4$, the integral over $\tau$ leads to a $\delta$-function establishing energy conservation $\delta\left[\omega(p) + \omega(q_1) + \omega(q_2) + \omega(q_3)\right]$. This $\delta$-function appears in the limit $t - t_0 \to \infty$. On mass-shell the arguments of the $\delta$-functions, establishing the energy and momentum conservation, cannot be simultaneously zero. Hence, in the absence of a background field, i.e. in a stationary situation, energy-momentum conservation forbids the change of level population and of the ground state. Then the ground state $\left| ground \right\rangle$ is actually a true vacuum state, $\left| vac \right\rangle$, of the free theory in question.  Which agrees with the discussion before equation (\ref{43}).

However, in the presence of a background field there is no energy conservation (or the energy is not bounded from below), because the system is not closed. In equation (\ref{aaaa}) this fact reveals itself via the presence of the exact harmonics $g_p(t)$, which are drastically different from the simple exponential. Therefore, the expressions for the level population and for the anomalous average are not zero at loop order.

Furthermore, in the presence of a background field the harmonics are functions of physical momenta. For example, in the constant electric field, $E$, background harmonics are functions of $k + eEt$ and, hence,
are invariant under the simultaneous compensating shifts of $k$ and $t$ (see e.g. \cite{Akhmedov:2014doa}, \cite{Akhmedov:2014hfa} for the detailed discussion). In de Sitter space, the physical momentum is $k \, e^{- Ht}$, where $H$ is the Hubble constant. Hence, the harmonics are invariant under compensating shift of time and rescaling of the momentum (see e.g. \cite{Akhmedov:2013vka} for the detailed discussion). Finally, in the collapse background at future infinity and in the vicinity of the horizon, harmonics are functions of $\omega \, e^{-t/r_g}$, where $\omega$ is the frequency and $r_g$ is the corresponding Schwarzschild radius of the collapsing body, as we will see in section \ref{harmiii}.
Because of these symmetries, after the change of integration variables from $t_3$ and $t_4$ to $T = (t_3 + t_4)/2$ and $\tau = t_3 - t_4$ in equation (\ref{aaaa}), one can get rid of the dependence of the integrand on $T$ for low enough external physical momenta. In such a case one finds that $\left\langle a^{\dagger}_{\vec{k}} \, a_{\vec{k}}\right\rangle(t) \propto \lambda^2 \, (t-t_0)$ and $\left\langle a_{\vec{k}} \, a_{-\vec{k}}\right\rangle(t) \propto \lambda^2 \, (t-t_0)$, where $(t-t_0)$ comes from the integration over $T$.
The coefficient of proportionality here can be interpreted as a part of the collision integral that is responsible for the change of the population numbers from the background field and its backreaction on the initial ground state. One important complication with respect to the standard condensed matter theory situations is the presence of the anomalous quantum average. In standard situations (see e.g. \cite{LL} and \cite{Kamenev}) the ground state of the theory does not change. In the latter cases the only quantity that receives growing corrections, if the initial state is chosen to be
non–stationary, is the particle number density and the formulae are a bit different from those in (\ref{aaaa}).

In the previous paragraph, we have assumed that the background field is always on---from the past to the future infinity. However, if the background field is turned on at some time $t_*$, then we can put $t_0$ before $t_*$. In such a situation, $\left\langle a^{\dagger}_{\vec{k}} \, a_{\vec{k}}\right\rangle(t) \propto \lambda^2 \, (t - t_*)$ and $\left\langle a_{\vec{k}} \, a_{-\vec{k}}\right\rangle(t) \propto \lambda^2 \, (t-t_*)$. In either case we have the secular growth of the loop corrections. Thus, after a long enough time of evolution we obtain $\lambda^2 \, t \sim 1$
and loop corrections become of the order of tree-level classical contributions. The corrections, thus, have to be resummed in all loops.~\footnote{Note that this is a signal of the breakdown of perturbations theory, which has a clear physical meaning. The resummation which has to be done is the standard procedure in condensed matter theory non--stationary situations \cite{LL}, \cite{Kamenev}. Apart from summation of the leading loop IR contributions it allows us to define the correct time evolution of the particle population numbers. In our case the main technical and conceptual complication is the presence of the anomalous quantum average, which makes this example similar to de Sitter space quantum field theory \cite{Akhmedov:2013vka}.}  Such a resummation has been done in the case of the constant electric field backgrounds in \cite{Akhmedov:2014doa}, \cite{Akhmedov:2014hfa} and in the case of de Sitter space massive scalar field theory in \cite{Akhmedov:2011pj}, \cite{Akhmedov:2012pa}, \cite{Akhmedov:2012dn}, \cite{Akhmedov:2013vka}, \cite{Akhmedov:2013xka} (see also \cite{Krotov:2010ma}, \cite{Polyakov:2012uc} and \cite{Serreau:2013psa}, \cite{Gautier:2013aoa}). In the present paper we start the same kind of study in the case of black hole radiation.

\section{The background geometry}
\label{sec:backgeo}

In this paper, we consider a spherically-symmetric, massive, thin shell that is kept fixed at radius $r = R_0$ (by an additional force) until $t=0$ when it is released and collapses in free-fall---see figure \ref{fig1}. Because of spherical symmetry, by  Birkhoff's theorem, the geometry is given by the Schwarzschild metric outside the shell and flat space inside:
\bqa\label{metric}
ds^2 = \left\{\begin{matrix}
dt_-^2 - dr^2 - r^2d\Omega^2, \quad & r  \leq R(t) & \\[2mm]
\left(1 - \frac{r_g}{r}\right) dt^2 - \frac{dr^2}{1 - \frac{r_g}{r}} -  r^2d\Omega^2, \, & r \geq R(t) &
\end{matrix}\right., \quad d\Omega^2 = d\theta^2 + \cos^2\theta d\varphi^2,
\eqa
where $R(t)$ is the radial coordinate of the shell, which before the start of collapse is $R(t \leq 0) = R_0$; $r_g/2$ is the ADM mass of the shell and $t$ ($t_-$) is the time coordinate outside (inside) the shell. We assume that $R_0 > r_g$ and, before the start of collapse, the shell is close to its Schwarzschild radius, i.e.  $|R_0 - r_g| \ll r_g$.

Note that, ideally, we would like to consider the collapse of some compact stellar object such as say a neutron star, whose radius is much larger than its Schwarzschild radius. However, here we assume that $|R_0 - r_g| \ll r_g$. In fact, in general, it is not possible for a stellar object of radius smaller than $3r_g/2$ to be stable. We consider such a phenomenologically unrealistic simplification, however, to make analytic headway. In particular, such a simplification allows us to find the behavior of the harmonic functions on such a background.

The background that we consider is described by two metrics (\ref{metric}) that have to be matched at the position of the shell \cite{Israel:1966rt, Israel:1967zz}. Before collapse, this is easily done because the radius of the shell is time-independent, i.e. on the shell $r=R_0$ and $dr=0$, and we have that
\bqa\label{rel1}
t_- = \sqrt{1-\frac{r_g}{R_0}} \, \, t, \quad t \leq 0.
\eqa
After the start of collapse, we assume that the shell is in the free fall, hence the metric on the shell is $ds^2 = d\tau^2 - R^2(\tau) \, d\Omega^2.$ Comparing this to the metric inside the shell, $$\left(\frac{dt_-}{d\tau}\right)^2 - \left(\frac{dR}{d\tau}\right)^2 = 1.$$ Furthermore, comparing the metric on the shell to the metric outside the shell gives a relation between $t$ and proper time of the shell $\tau$,
\bqa
\left(1 - \frac{r_g}{R}\right)\, \left(\frac{dt}{d\tau}\right)^2 - \frac{\dot{R}^2}{1 - \frac{r_g}{R}} = 1,
\eqa
where $\dot{R} = \frac{dR}{d\tau}$. Using our assumption that $|R-r_g| \ll r_g$, noting that $\dot{R} \neq 0$ and $\frac{dt}{d\tau} \to \infty$ as $t \to \infty$, we neglect the right handside in comparison to the left handside  and integrate the above relation to find the shell's trajectory from the point of view of the outside observer:
\bqa \label{Rinf}
R(t) \approx r_g\left(1 + \frac{R_0-r_g}{r_g} \, e^{-\frac{t}{r_g}}\right).
\eqa
From the point of view of the inside observer, given our assumption that  $\left|R_0 - r_g\right| \ll r_g$, the shell collapses at constant speed $\nu$ \cite{Israel:1967zz}, where
\bqa \label{Rtminf}
\nu\equiv \left|\frac{dR(t_-)}{dt_-}\right| = \frac{\left|\dot{R}\right|}{\sqrt{1 + \dot{R}^2}}, \quad {\rm and} \quad R(t_-) \approx R_0 - \nu t_-.
\eqa
Now, identifying $R(t) = R(t_-)$ and using equations (\ref{Rinf}) and (\ref{Rtminf}), we find the relation between $t_-$ and $t$:
\bqa\label{rel2}
t_-  \approx \frac{R_0 - r_g}{\nu}\left(1 - e^{-\frac{t}{r_g}}\right), \quad t \to \infty.
\eqa
As $t\to \infty$, the shell approaches its Schwarzschild radius $R(t) \to r_g$. On the other hand, $t_- \approx \left(R_0 - r_g\right)/\nu < \infty$ is the moment of internal time when the horizon is formed. Beyond this moment of time there is no relation between $t$ and $t_-$.
In section \ref{sec:har}, we use the relations between $t$ and $t_-$ before the collapse and at its final stage, relations (\ref{rel1}) and (\ref{rel2}) respectively, to find the behavior of the free scalar harmonics in these regimes.

In deriving the behavior of the harmonics during collapse, we also use another simplifying assumption that we will now explain. Gluing the two metrics in (\ref{metric}) to each other with the use of the shell's energy-momentum tensor, we obtain the relation \cite{Israel:1966rt} (see also \cite{Poisson}):
\bqa\label{energy}
\frac{r_g}{2} = M\sqrt{1 + \dot{R}^2} - \frac{M^2}{2 R},
\eqa
where $M=const$ defines the energy-momentum tensor of the shell's matter content. This relation expresses the fact that the total energy of the shell, $r_g/2$, is the sum of the kinetic energy, $M\sqrt{1+\dot{R}^2}$ (recall that $\tau$ is proper time rather than coordinate time), and its potential energy, $M^2/2R$. To avoid a bounce, when $\dot{R} = 0$, i.e. to have the actual collapse, we must have that $r_g > 2M$. However, we in fact make a stronger assumption, namely that $r_g \gg 2M$. Neglecting the difference between $R(\tau)$ and $r_g$, from equation \eqref{energy} we obtain

$$\dot{R} \approx - \sqrt{\left(\frac{r_g}{2M} + \frac{M}{2r_g}\right)^2 - 1}.$$
Hence our assumption that $r_g \gg 2M$ implies that  $\left|\dot{R}\right| \gg 1$ and $\nu \approx 1$, i.e. the shell is almost light-like at the final stage of the collapse even from the point of view of the inside observer. While we use $\nu \approx 1$, we keep $\nu$ explicitly in most of the formulae presented below to indicate how the situation would be different without this assumption.

\begin{figure}
\begin{center}
\includegraphics[scale=0.3]{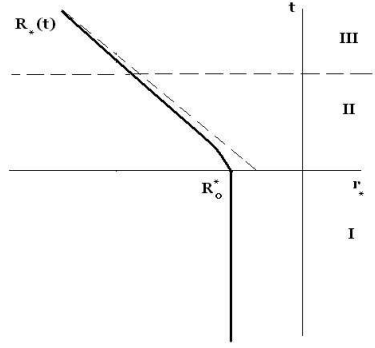}\caption{The collapse of a massive, thin shell in tortoise coordinates. We again divide the collapse into three phases: I, II and III}\label{fig2}
\end{center}
\end{figure}

In the region outside the shell, it is convenient to also use tortoise coordinates
$$r_* = r + r_g \log\left(\frac{r}{r_g} - 1\right).$$
In terms of these coordinates the trajectory of the shell at the final stage of the collapse is
\bqa \label{R*inf}
R_*(t) \approx R^*_0 - t + (r_g -R_0) \left(1 - e^{-\frac{t}{r_g}}\right),
\eqa
where $R_0^* = R_0 + r_g \log \left(\frac{R_0}{r_g} - 1\right)$; i.e. the shell's motion becomes light-like as $t\to \infty$---see figure \ref{fig2}.

\section{Free harmonics} \label{sec:har}

The theory that we study on the background of a collapsing shell is the real scalar field theory with $\phi^4$ interaction:
\bqa
S = \int d^4 x \sqrt{|g|}\left[ (\partial_\mu \phi)^2 - m^2 \phi^2 - \frac{\lambda}{4!} \phi^4\right].
\eqa
In this section, we discuss the free theory case $\lambda = 0$.

We expand the field $\phi$ as follows:
$$
\phi\left(t,r,\theta,\varphi\right) = \sum_{l,n} Y_{l,n}\left(\theta,\varphi\right)\, \phi_{l}\left(t,r\right),
$$
where $Y^*_{l,n}\left(\theta,\varphi\right) = Y_{l,n}\left(\theta,\varphi\right)$ are real spherical harmonics. Given this expansion and the background given by equation \eqref{metric}, the action takes the form
\begin{align}\label{confac}
S &= \sum_{l} \left(2l + 1\right) \, \int dt \int^{R(t)}_0 dr \, r^2 \, \left(\frac{\partial t_-}{\partial t}\right) \, \left[\left(\frac{\partial t}{\partial t_-}\right)^2 \left(\partial_{t} \phi_{l}\right)^2 - \left(\partial_r \phi_{l}\right)^2 - \left(\frac{l(l+1)}{r^2} + m^2\right) \phi_{l}^2 \right] \notag \\[2mm]
& \qquad + \sum_{l} \left(2l + 1\right) \, \int dt \int^\infty_{R(t)} dr \, r^2\,  \left[ \frac{\left(\partial_{t} \phi_{l}\right)^2}{1 - \frac{r_g}{r}} - \left(1 - \frac{r_g}{r}\right) \, \left(\partial_r \phi_{l}\right)^2 - \left(\frac{l(l+1)}{r^2} + m^2\right) \phi_{l}^2 \right].
\end{align}
Varying this action, we obtain the equations of motion,
\bqa\label{eoms}
\begin{cases}
\left[\partial_{t_-}^2 - \partial_r^2 + m^2 + \frac{l(l+1)}{r^2}\right] \, \left(r \phi_{l}\right) = 0, & \quad r \leq R(t) \\[2mm]
\left[\partial_{t}^2 - \partial_{r^*}^2 + \left(1 - \frac{r_g}{r}\right)\left(m ^2 + \frac{l(l+1)}{r^2} + \frac{r_g}{r^3}\right)\right] \, \left(r \phi_{l}\right) = 0, & \quad r \geq R(t)
\end{cases},
\eqa
and the boundary conditions,
\begin{align}\label{bound}
\phi_{l}\Bigl[R(t) -  0 \Bigr] &= \phi_{l}\Bigl[R(t) +  0\Bigr], \notag\\[4mm]
\left[\left(\frac{\partial t}{\partial t_-}\right)\, \left|\frac{dR}{dt}\right|\, \partial_{t} \phi_{l}   - \left(\frac{\partial t_-}{\partial t} \right) \partial_r \phi_{l}\right]_{r=R(t) - 0} &=
\left[\frac{\partial_{t} \phi_{l}}{1 - \frac{r_g}{r}} \left|\frac{dR}{dt}\right|  - \left(1 - \frac{r_g}{r}\right) \, \partial_r \phi_{l}\right]_{r=R(t) + 0}.
\end{align}
The second equation above relates the normal derivative of the scalar field across the shell. As $t \to \infty$, the derivative normal to the shell outside it becomes the derivative with respect to the retarded coordinate $u= t- r_{*}$.

The equations of motion outside the shell can be understood in terms of confluent Heun equations \cite{Leaver, Fiziev:2005ki, Philipp:2015jja} and  there has been recent progress in understanding the properties of solutions to these equations \cite{Ronveaux, Slavyanov}. However, for our purposes it is sufficient to consider approximate solutions of these equations in various \underline{stationary} regions of space-time. In this way we obtain estimates of the leading effects/contributions to Hawking radiation.

The reason we can find approximate solutions to equations (\ref{eoms}) and (\ref{bound}) lies in the fact that in the stationary situation we can separate the variables $t$ and $r$. Furthermore, the potential,
\bqa\label{poten}
U(r) = \left(1 - \frac{r_g}{r}\right)\left[m ^2 + \frac{l(l+1)}{r^2} + \frac{r_g}{r^3}\right], \quad r > R_0,
\eqa
in the second equation of (\ref{eoms}) is vanishing as $r\to r_g$ and approaches a constant, $m^2$, as $r\to \infty$. Therefore, in the vicinity of the shell---note the relation $\left|R(t) - r_g\right| \ll r_g$---the harmonics can be approximated by plane waves.
Meanwhile, at asymptotic spatial infinity, $r\to \infty$, we have that $r\approx r_*$ and the harmonics are again plane waves. The relation between the harmonics in the vicinity of the shell and at spatial infinity can be found by solving the scattering problem across the potential barrier (\ref{poten}). In the next subsection, we will use the above arguments to find the approximate harmonics.

\begin{figure}
\begin{center}
\includegraphics[scale=0.4]{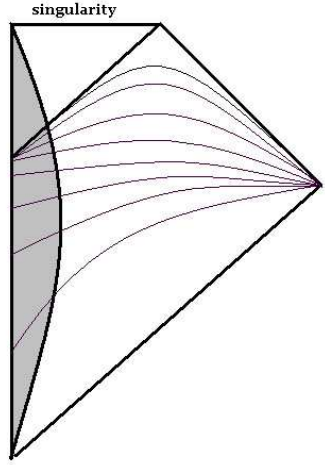}\caption{Penrose diagram of the collapsing shell background. The curved, thin lines depict Cauchy surfaces with respect to the Schwarzschild time $t$. The grey region represents the shell interior.}\label{fig3}
\end{center}
\end{figure}

Using the equations of motion, the free Hamiltonian of the massive scalar theory can be rewritten as
\bqa \label{hamstat}
H_0(t) = \sum_{l} (2l+1)\, \int^{\infty}_0 dr \, \frac{\sqrt{|g|}}{\sin(\theta)} \, \left[g^{tt} \left(\partial_t \phi_{l}\right)^{2} - \frac{1}{\sqrt{|g|}} \phi_{l} \partial_{t} \left( \sqrt{|g|} g^{tt} \partial_t \phi_{l} \right) \right].
\eqa
In the collapsing shell background, using the metric given in equation (\ref{metric}), this Hamiltonian is
\begin{align} \label{ham}
H_0(t) &= \sum_{l,n} \int\limits^{R(t)}_0 r^2 dr  \left[\left(\frac{\partial t}{\partial t_-}\right) \left(\partial_{t} \phi_{l}\right)^2 - \phi_{l} \, \partial_{t} \left( \left(\frac{\partial t}{\partial t_-}\right) \partial_{t} \phi_{l} \right)\right] +
\sum_{l,n} \int\limits^{\infty}_{R(t)}  \frac{r^2 \,dr}{1 - \frac{r_g}{r}} \left[\left(\partial_{t} \phi_{l}\right)^2 - \phi_{l} \partial^2_{t} \phi_{l} \right] \nonumber \\[3mm]
& \qquad + \, \sum_{l,n} R^2(t) \, \phi_{l}\left[\left. \left(\frac{\partial t_-}{\partial t}\right) \partial_r \phi_{l} \, \right|_{r = R(t)-0} - \left. \left(1 - \frac{r_g}{R(t)}\right) \, \partial_r \phi_{l} \, \right|_{r = R(t)+0}\right].
\end{align}
The last term in this expression is the contribution of the field values on the shell itself.
This Hamiltonian defines translations along the time coordinate $t$. The corresponding Cauchy surfaces are depicted in the figure \ref{fig3}.

\subsection{In-harmonics before collapse}

During the first stage, when the shell is stationary and is yet to collapse, we can find harmonics that diagonalize the free hamiltonian, hence providing a sensible definition of particle number. The state with respect to which the flux is found is defined in terms of these harmonics, which we call in-harmonics.

As is usual in canonical quantization, we expand the scalar field in terms of a basis of harmonics $\bar{h}^{\phantom{\frac12}}_{\omega, l}(r,t)$:
\bqa\label{expan}
\phi\left(\underline{x}, t\right) =  \sum_{l, n} Y_{l, n} (\theta, \varphi) \int_{m}^{\infty} \frac{d\omega}{2\pi} \left[ a_{\omega, l, n} \bar{h}^{\phantom{\frac12}}_{\omega, l}(r,t) + {\rm h.c.} \right], \nonumber \\
\pi\left(\underline{x}, t\right) = g^{tt}  \sum_{l, n} Y_{l, n} (\theta, \varphi) \int_{m}^{\infty} \frac{d\omega}{2\pi} \,  \left[ a_{\omega, l, n} \partial_{t} \bar{h}^{\phantom{\frac12}}_{\omega, l}(r,t) + {\rm h.c.} \right],
\eqa
where, at this stage, $\omega$ simply labels the harmonics and ``${\rm h.c.}$'' stands for hermitian conjugate. The harmonics $\bar{h}^{\phantom{\frac12}}_{\omega, l}(r,t)$ provide a basis for solutions of the Klein-Gordon equation (\ref{eoms}). Note that we only expand in harmonics with $\omega >m$ because only these modes are oscillatory at spatial infinity. We explain this point further below equation \eqref{modeos}.
The harmonic functions that we wish to work with are such that they diagonalize the Hamiltonian when the shell is stationary. We refer to the corresponding ground state, which is annihilated by all annihilation operators, $a_{\omega, l, n} |in\rangle = 0$, as the in-state.

The canonical commutation relations for the field $\phi\left(\underline{x}, t\right)$ are
\bqa \label{cancom}
[\phi(\underline{x}, t) , \pi(\underline{y}, t)] = i \delta^{(3)}\left(\underline{x} - \underline{y}\right)
\eqa
and all other commutations relations vanish. Assuming,
\bqa
[a^{\phantom{\dagger}}_{\omega, l, n}, a^{\dagger}_{\omega', l', n'}] = 2 \pi \delta_{l l'} \, \delta_{n n'} \, \delta(\omega - \omega'),
\eqa
we obtain the following condition on the modes from the canonical commutation relations
\bqa \label{cancomcon}
\sum_{l,n} Y_{l, n} (\theta, \varphi) Y_{l, n} (\theta', \varphi') g^{tt} \, \int_{m}^{\infty} \frac{d\omega}{2 \pi} \, \left[\bar{h}_{\omega, l}(t, r)\,  \partial_{t} \bar{h}^{*\phantom{\frac12}}_{\omega, l}(t, r') - {\rm h.c.} \right] = i \, \delta^{(3)}\left(\underline{x} - \underline{x}'\right).
\eqa
We stress that $\underline{x}$ are the spatial coordinates.

In a stationary situation we can separate the dependence of $\bar{h}_{\omega, l}(t,r)$ on $t$ and $r$.
Then during the stage before collapse, the positive energy states can be represented as
$$\bar{h}_{\omega, l}(t,r) = h_{\omega, l}(r) \, e^{-i \omega t} = h_{\omega, l}(r) \, e^{- i \omega_- t_-},$$
where from equation (\ref{rel1}) it follows that $\omega_- = \omega/\sqrt{1 - \frac{r_g}{R_0}}$. Given the decomposition above, we can identify the label $\omega$ in equation (\ref{expan}) with the energy of the mode $\bar{h}_{\omega, t}(t,r)$ at past infinity.

Inside the shell, by solving the first equation in \eqref{eoms}, we find that
\bqa \label{modeis}
\bar{h}_{\omega, l}(t,r) = \frac{\A_{\omega}}{\sqrt{r}} J_{l+\frac{1}{2}} \left(\sqrt{\omega_-^2 -m^2} \; r\right) \, e^{- i \, \omega_- \, t_-}, \quad {\rm for} \quad r \leq R_{0} \quad {\rm and} \quad t \leq 0.
\eqa
Note that we only require Bessel functions of the first kind, which lead to finite harmonics at $r=0$.~\footnote{Note that these Bessel functions lead to the diagonalization of the free Hamiltonian in empty flat space-time in spherical coordinates.}  From equation \eqref{modeis}, we can see that $\omega$ is bounded from below: $\omega_- \geq m$ or $\omega \geq m_- \equiv m \, \sqrt{1 - r_g/R_0}$. If the mass of the scalar is small, ${\cal A}_\omega$ can be approximated by $\sqrt{\pi}$---see appendix \ref{app:norm}. Note that it is not necessary to make this assumption and we could simply keep ${\cal A}_\omega$ in the expressions without solving the normalization condition \eqref{cancomcon}. However, we will use  ${\cal A}_\omega \approx \sqrt{\pi}$ for convenience.

Similarly, from the second equation in (\ref{eoms}), we find that, before $t=0$, the harmonics outside the shell are given by
\bqa \label{modeos}
\bar{h}_{\omega, l}(t,r) = \frac{e^{-i \, \omega \, t}}{r}
\begin{cases}
A_{\omega} e^{-i \omega r_{*}} + B_{\omega} e^{i \omega r_{*}}, \qquad \quad |r - R_0| \ll r_g, \\
C_\omega \, e^{- i k r_*} + D_\omega \, e^{i k r_*},  \quad \quad \quad r \gg R_0,
\end{cases}
\eqa
where $ k = \sqrt{\omega^2 - m^2}$.

Now we can see that when $\omega \geq m$, the harmonics oscillate as $r_* \to \infty$. Hence, in such a case we have the continuous spectrum. But when $m_- \leq \omega \leq m$ the harmonics decay or grow exponentially as $r_* \to \infty$. After dropping the exponentially growing mode, we are left with modes that correspond to a discrete spectrum that oscillate between $r=0$ and the turning point of the potential (\ref{poten}). From the gluing conditions across the potential (\ref{poten}) one can estimate the allowed values of the discrete energy levels $\omega_i$. We leave the study of these discrete modes for the future. For our present purposes, it is important to know that there are only finitely many such states in the well, which has a depth of order $m^2 \, r_g/R_0$. Hence, these states play a minor role in the effects that we consider here. In particular, since they do not correspond to running states at spatial infinity, they do not contribute to Hawking radiation at tree-level. Moreover, if we take $m^2 \, r_g/R_0$ to be small enough---we are in fact working with small $m$---there will be no discrete states at all.

The absolute value of the undetermined coefficients in equation \eqref{modeos} can be found using the normalization condition \eqref{cancomcon}. However, since we are only interested in the behavior of the harmonics in the vicinity of the shell, we will concentrate on coefficients $A_{\omega}$ and $B_{\omega}$. The effects that we are looking for appear in the vicinity of the collapsing shell. As seen from spatial infinity, or elsewhere, these effects simply receive grey body factors due to the potential barrier (\ref{poten}).

We find coefficients $A_{\omega}$ and $B_{\omega}$ using the boundary conditions \eqref{bound}.  During the first stage, when $R(t) = R_0$, the boundary conditions reduce to
\bqa \label{statbound}
h_{\omega, l} \Bigl[R_0 -  0 \Bigr] = h_{\omega, l} \Bigl[R_0 + 0 \Bigr], \qquad
\left[ \partial_r h_{\omega, l}\right]_{r=R_0 - 0} =
\sqrt{1 - \frac{r_g}{R_0}} \, \left[  \partial_r h_{\omega, l}\right]_{r=R_0 + 0}.
\eqa
In appendix \ref{app:ABcoeff}, we show that the above equations imply that
\bqa \label{coeffmodeso}
A_\omega = B_{\omega}^* = \frac{i^{l+1}}{ \sqrt{2\, \omega}} \left( 1- \frac{r_g}{R_0} \right)^{1/4}  e^{i \, \omega \, \left[R_0^* - R_0\, \left(1-\frac{r_g}{R_0}\right)^{-\frac12} \right] } + \mathcal{O}\left(\left( 1- \frac{r_g}{R_0} \right)^{3/4} \right).
\eqa

Let us consider the free Hamiltonian for these modes (see reference \cite{GribMamaevMostepanenko} for a similar discussion). In a generic situation, the free Hamiltonian, expressed via harmonic functions, takes the following form:
\bqa
H_0(t) = \sum_{l,n} \iint_{m}^{\infty} \frac{d\omega d\omega'}{(2\pi)^2} \, \left[{\cal E}_{\omega,\omega',l}(t) a^{\dagger}_{\omega,l,n}\, a_{\omega',l,n} + {\cal J}_{\omega, \omega',l}(t) a_{\omega,l,n}\, a_{\omega',l,n} + {\rm h.c.}\right],
\eqa
where
\begin{align*}
{\cal E}_{\omega,\omega',l}(t) &= \int_0^{\infty} dr \frac{\sqrt{|g|}}{\sin(\theta)}\left\{g^{tt} \partial_t \bar{h}^*_{\omega,l}(r,t)\, \partial_t \bar{h}_{\omega',l}(r,t) - \frac{1}{\sqrt{|g|}} \bar{h}^*_{\omega, l}(r,t) \partial_t \left[\sqrt{|g|} \, g^{tt}\partial_t \bar{h}_{\omega',l}(r,t)\right]\right\}, \nonumber \\[2mm]
{\cal J}_{\omega,\omega',l}(t) &= \int_0^{\infty} dr \frac{\sqrt{|g|}}{\sin(\theta)}\left\{g^{tt} \partial_t \bar{h}_{\omega,l}(r,t)\,\partial_t \bar{h}_{\omega',l}(r,t) - \frac{1}{\sqrt{|g|}} \bar{h}_{\omega, l}(r,t) \partial_t \left[\sqrt{|g|} \, g^{tt}\partial_t \bar{h}_{\omega',l}(r,t)\right]\right\}.
\end{align*}
Thus, generically the free Hamiltonian is not diagonal because of the presence of terms with $\omega \neq \omega'$ and especially because of the presence of a non-zero ${\cal J}_{\omega, \omega', l}$.
However, given suitably chosen normalizable modes, in stationary situations ${\cal E}$ and ${\cal J}$ become constants because of the factorization of the dependence of the harmonic functions on $t$ and $r$: $\bar{h}_{\omega, l}(r,t) = e^{- i \omega t}\, h_{\omega, l}(r)$. In such a case one can find harmonic functions, satisfying
\bqa \label{modenorm}
\int_{0}^{\infty} dr \, \frac{\sqrt{|g|}}{\sin\theta}\, g^{tt} \, h_{\omega,l}(r) \, h_{\omega',l}(r) = 0, \qquad \quad \int_{0}^{\infty} dr \, \frac{\sqrt{|g|}}{\sin\theta} \, g^{tt} \, h_{\omega,l}(r) \, h_{\omega',l}^*(r) = \frac{\pi}{\omega} \delta(\omega - \omega'),
\eqa
which lead to the vanishing ${\cal J}$ terms. Note that the normalization above is chosen so as to be consistent with condition \eqref{cancomcon} in stationary situations.

The stage before collapse, considered in this section, is indeed a stationary situation, but we have discussed the behavior of the $h_{\omega, l}(r)$ modes for certain ranges of $r$. To be able to find the free Hamiltonian, we integrate over $r$ and hence, in principle, we need the modes for all $r$.  However, here it is sufficient to assume that there are normalizable modes that satisfy equation \eqref{modenorm}---these modes will necessarily have the form given in equations \eqref{modeis} and \eqref{modeos} for the appropriate regions. With the normalization of the modes given in equation \eqref{modenorm}, it is simple to verify that the free Hamiltonian is
\bqa
H_0(t\leq 0) = \sum_{l, n} \int_{m}^{\infty} \frac{d \omega}{2\pi} \omega \left[ a^{\phantom{\dagger}}_{\omega, l, n} a^{\dagger}_{\omega, l, n} + a^{\dagger}_{\omega, l, n} a^{\phantom{\dagger}}_{\omega, l, n} \right]
\eqa
  and is indeed diagonal before the start of the collapse. But during collapse, the in-harmonics no longer diagonalize the Hamiltonian. This is a sign that there is particle creation.

\subsection{In-harmonics during the late-stage of collapse} \label{harmiii}

\begin{figure}
\begin{center}
\includegraphics[scale=0.3]{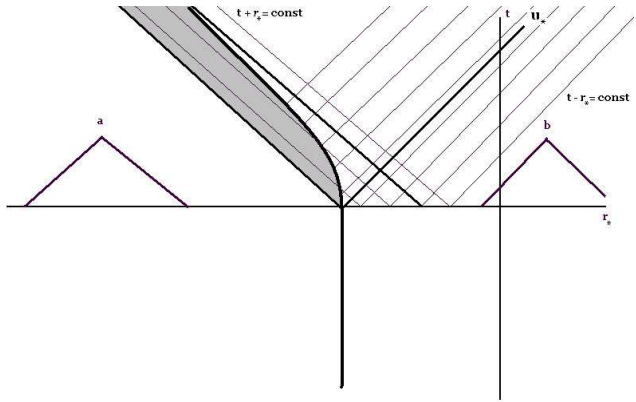}\caption{An illustration of the Cauchy problem for harmonics after collapse.}\label{fig4}
\end{center}
\end{figure}

In this section, we find the behavior of in-harmonics as $t \longrightarrow \infty$. The change in the behavior of the in-harmonics from past to future infinity due to the non-stationarity of the background is what causes particle creation.

We assume that $\nu \approx 1$ and neglect the difference between $\nu$ and 1. The reason for this is as follows: Given a harmonic function inside the shell, see figure \ref{fig4}, we would like to find the harmonic function outside the shell by solving the gluing condition \eqref{bound}. To find the form of the harmonic function behind the shell one has to solve the Cauchy problem with initial values given at $t=0=t_-$. The initial condition is given by the values of the harmonic functions behind the shell and just outside it and their derivatives at $t=0=t_-$.
Then, at a point inside the shell which is causally disconnected from the region outside of the shell at $t=0$, e.g.\ a point labeled ``$a$'' in figure (\ref{fig4}), the value of the harmonic is simply given by equation (\ref{modeis}). However, in order to use the boundary condition to find the harmonic functions outside the shell, we need to solve the Cauchy problem for the region inside the shell, grey region in figure \ref{fig4}, that is within the domain of influence of points outside the shell on the initial Cauchy hypersurface. This appears to be a rather complicated problem. However, if we take $\nu \approx 1$ then the domain of dependence of points inside the shell are always inside the shell, hence the grey region can be neglected and the matching can be done straightforwardly, as we will see below.

Assuming $\nu \approx 1$, the harmonics inside the shell are given by equation (\ref{modeis}), where $\omega_- = \omega/\sqrt{1 - r_g/R_0}$. Since we are only interested in the modes near the exterior of the shell, assuming $|R(t) - r_g| \ll r_g$, we neglect the potential \eqref{poten}. The solution is again as in the case before collapse and can be given in terms of Fourier modes. However, as the matching at the shell is more complicated with the solution in this form, we instead keep the solution general. In all we have the following solution, as $t\to +\infty$,
\bqa \label{modeiii}
\bar{h}_{\omega, l}(t, r) = \frac{1}{r}
\begin{cases}
\sqrt{\pi \, r} J_{l+\frac{1}{2}} \left(\sqrt{\omega_-^2 -m^2} \; r\right) e^{\pm i \omega_- t_-}, \quad r \leq R(t), \\
f_{\omega, l}(u) + g_{\omega, l} (v), \hspace{33.5mm} r\geq R(t), \; |r - R(t)| \ll r_g,
\end{cases}
\eqa
where $u= t - r_{*}$ and $ v= t+ r_*$ are outgoing and ingoing null coordinates, respectively. Since the shell is light-like at the late-stage of collapse, the $v$-dependent part of the harmonics outside the shell is not affected by the collapse. This can be easily understood by noting that the boundary conditions are continuity of the harmonics and their  derivatives normal to shell. Hence, $g_{\omega,l}(v)$ is unaffected by the collapse and, from equation \eqref{modeos}, we have
\begin{equation} \label{giii}
g_{\omega, l} = A_\omega \, e^{-i \, \omega \, v}.
\end{equation}
We now find the $u$-dependent part of the harmonic outside the shell by imposing the boundary conditions \eqref{bound} at the shell. Note that we are assuming that $\nu \approx 1$, which means that the shell collapses at approximately the speed of light even from the point of view of observers inside the shell. In terms of the picture in figure \ref{fig4}, this means that we are assuming that the grey region is infinitesimally thin. In this approximation the harmonics inside the shell are given by Bessel functions, as in the static case, but close to the shell they also have a constant contribution that is equal to the value of the $v$-dependent part of the harmonics outside the shell. Therefore, taking this constant piece into account, for the positive frequency modes the continuity of $\phi_l$ across the shell gives
\bqa\label{cond1}
\sqrt{\pi \, R} J_{l+\frac{1}{2}} \left(\sqrt{\omega_-^2 -m^2} \; R\right) e^{- i \omega_- t_-} = \left[ f_{\omega, l}(u) \right]_{r = R}.
\eqa
For brevity we denote $R(t)$ as $R$, but the dependence of $R$ on $t$ should be remembered. Using equation \eqref{R*inf}, as $ t \longrightarrow \infty$ the outgoing null coordinates evaluated at the shell are approximately
\bqa \label{uvapp}
\left[ u \right]_{r = R}  \approx 2t -(R^{*}_0 + r_g - R_0).
\eqa
Hence, the continuity relation reduces to
\bqa
f_{\omega, l}\left[2t -\left(R^{*}_0 + r_g - R_0\right)\right] \approx \sqrt{\pi \, R} J_{l+\frac{1}{2}} \left(\sqrt{\omega_-^2 -m^2} \; R\right) e^{- i \omega_- t_-}.
\eqa
Therefore, using relation \eqref{rel2}, we find that
\bqa\label{f1}
r \, \bar{h}_{\omega, l}(r,t) \approx \sqrt{\pi \, R(u)} J_{l+\frac{1}{2}} \left[\sqrt{\omega_-^2 -m^2} \; R(u) \right] e^{\; - i \omega_- \frac{(R_0 -r_g)}{\nu} \left( 1- e^{-\frac{u+ R_0^* +r_g - R_0}{2 r_g}}\right)} + g_{\omega, l} (v),
\eqa
where $g_{\omega, l}(v)$ is given in equation \eqref{giii} and
\bqa \label{Ru}
R(u) = r_g \left( 1+ \frac{R_0-r_g}{r_g} e^{-\frac{u+ R_0^* +r_g - R_0}{2 r_g}} \right).
\eqa
As $t \longrightarrow \infty$, the boundary condition for the derivative of the modes at the shell, \eqref{bound}, becomes
\bqa \label{dbconinf}
\left(\frac{\partial t_-}{\partial t} \right) \left[\nu \, \partial_{t_-} h_{\omega, l}   -  \partial_r h_{\omega, l}\right]_{r=R(t) - 0} = 2
\left[ \partial_{u} h_{\omega, l} \right]_{r=R(t) + 0}.
\eqa
Since, as $t \longrightarrow \infty$, $R \longrightarrow r_g,$ we have used $\frac{R}{r_g} \approx 1$ to simplify the right-handside of the above equation. In Appendix \ref{app:boundcon}, we show that this relation is satisfied by (\ref{f1})
if we neglect the difference between $\nu$ and 1. In the case where $\nu$ is much less than 1, our solution for the $v$-dependent part of the harmonic will be modified and will be more complicated.

It is worth stressing at this point that the behavior of the harmonic function very far from the gravitating center (at spatial infinity) is the same both when the shell is static and when the shell is collapsing.
That can be expected on general physical grounds because of locality. Hence, while the harmonics have dependence on null coordinates near the shell surface, this will not be the case at spatial infinity. This is expected because these modes are harmonics of a massive scalar field which propagates to future time--like infinity and not future null infinity.

\section{Hawking radiation (tree-level)} \label{sec:hawtree}

We will, first, reproduce the standard thermal radiation using the harmonics constructed in the previous section. In particular, we calculate the energy flux at the final stage of collapse. It is technically easier and physically more appropriate to calculate it just outside the shell, $r \gtrsim r_g$, and then continue across the potential barrier (\ref{poten}). In fact, the flux is created just outside the black hole. Thus, we find the flux as $t\to \infty$ in the vicinity of the shell, $|r - r_g| \ll r_g$.

In the proximity of the shell, the energy flux is given by
\bqa \label{flux1}
J\left(r \approx r_g, t\right) \equiv \int_{S_2} \sin\theta \, d\theta \, d\varphi \, r^2 \, \left\langle : {T^{r{\phantom{\frac12}}}}_{t}\left(r, t\right) : \right\rangle \approx - r^2_g  \, \int_{S_2} \sin \theta \,  d\theta \, d\varphi \, \left\langle : T^{\phantom{\frac12}}_{t r_*}\left(r, t\right) : \right\rangle.
\eqa
Using the expression for the energy-momentum tensor of a scalar field
\begin{equation} \label{flux2}
  J\left(r \approx r_g, t\right) = \sum_{l} \left(2l + 1\right) \left(J_{u}^{(l)} - J_{v}^{(l)} \right),
\end{equation}
where
\begin{align}
 J_{u}^{(l)} &= r_g^2 \int_{m}^{\infty} \frac{d\omega}{2\pi} \left[ \partial_u \bar{h}^{*}_{\omega,l}(r,t)^{\phantom{\frac12}} \partial_u \bar{h}_{\omega,l}(r,t) + {\rm c.c.}\right], \notag \\[2mm]
 J_{v}^{(l)} &= r_g^2 \int_{m}^{\infty} \frac{d\omega}{2\pi} \left[ \partial_v \bar{h}^{*}_{\omega,l}(r,t)^{\phantom{\frac12}} \partial_v \bar{h}_{\omega,l}(r,t) + {\rm c.c.}\right].
\end{align}
In equation \eqref{flux1}, the normal ordering is understood in the standard way in the presence of the background fields. In particular, we understand $\left\langle : T^\mu_\nu : \right\rangle$ as a subtraction from $\left\langle T^\mu_\nu \right\rangle$ of the same expression but calculated for the case of a shell that is eternally static at $r=R_0$. In this case, since the harmonic functions in the vicinity of a static shell are given by Fourier modes in $u$ and $v$, from equation \eqref{flux2}, the flux vanishes.

In order to find the flux due to the collapsing shell we use the harmonics outside the shell at the late-stage of collapse given in equation \eqref{f1},
\begin{align}\label{hIII}
\bar{h}_{\omega, l}(r,t) &\approx \frac{1}{r_g}\, \left(1 - \frac{r_g}{R_0}\right)^\frac14\, \sqrt{\frac{2}{\omega}} \, \cos\left[\frac{\pi \, (l+1)}{2} - \omega_- r_g\right] \, e^{i \, \omega r_g \, e^{-\frac{u - u_0}{2r_g}}}  \nonumber \\[2mm]
&\hspace{45mm} + \, \frac{1}{r_g} \, \left(1 - \frac{r_g}{R_0}\right)^\frac14\, \frac{i^{l+1}}{\sqrt{2\omega}} \, e^{-i \, \omega \, v + i \, \omega \, [R^*_{0} - R_0 (1-r_g/R_0)^{-1/2}]},
\end{align}
where $u_0 =  r_g \log(R_0/r_g) - r_g$. In order to obtain this expression from equation (\ref{f1}), we use $R(u)\approx r_g$, $\nu \approx 1$; neglect the difference between $r$ and $r_g$ in the denominator and neglect $m^2$ in comparison with $\omega_-^2 = \omega^2/\left(1 - r_g/R_0\right)$, because $R_0 \approx r_g$ and $\omega \geq m$ for the continuous part of the spectrum. We also use the limiting form of the Bessel function of the half-integer index with large argument.

In order to calculate the flux, we now simply need to insert the harmonic given in equation \eqref{hIII} into equation (\ref{flux2}) and evaluate the integrals. However, in order to make contact with the original derivation of Hawking radiation \cite{Hawking:1974sw} and for simplicity, it is appropriate to re-expand the expression on the right handside of equation (\ref{hIII}) in terms of the modes given in equation (\ref{modeos}). Because the $v$-dependent part is unchanged we only have to re-expand the $u$-dependent part:
\bqa \label{expanh}
\left(1 - \frac{r_g}{R_0}\right)^\frac14 \, \sqrt{\frac{2}{\omega}} \, \cos\left[\frac{\pi \, (l+1)}{2} - \omega_- r_g\right] \, e^{i \, \omega r_g \, e^{-\frac{u - u_0}{2r_g}}} =  \int_{|\omega'|>m} \frac{d\omega'}{2\pi\sqrt{2|\omega'|}} \, \alpha(\omega,\omega') \, e^{-i\omega' u},
\eqa
where $ \alpha_{\omega,\omega'} = \alpha(\omega, |\omega'|)$ and $\beta_{\omega,\omega'} = \alpha(\omega,-|\omega'|)$ are proportional to the seminal Bogoliubov coefficients for Hawking radiation.

The implicit expression for $\alpha(\omega, \omega')$ given in equation \eqref{expanh} is in fact only valid for $t>0$. There is also a contribution to $\alpha(\omega, \omega')$ from before collapse. However, these contributions will be of the form $\delta( \omega - \omega')$. Since we are only interested in the effects due to collapse, we will ignore these terms in $\alpha(\omega, \omega')$ and, from equation \eqref{expanh}, find
\bqa
\alpha\left(\omega,\omega'\right) \approx 2\, \left(1 - \frac{r_g}{R_0}\right)^\frac14 \, \sqrt{\frac{|\omega'|}{\omega}} \, \cos\left[\frac{\pi \, (l+1)}{2} - \omega_- r_g\right] \int_{u_*}^\infty du \, e^{i\omega r_g \, e^{-\frac{u-u_0}{2r_g}}} \, e^{i\omega' u},
\eqa
where $u_* = - R_0^*$. Because of the rapid oscillations of the integrand in the lower limit of the $u$ integration compared with the upper limit, the exact position of $u_*$ is not very relevant and we can extend the range of integration to the real line. Moreover, as we have eluded to before, since we are interested in the late-stage flux generated by the collapsing shell, we have ignored contributions to $\alpha\left(\omega,\omega'\right)$ from stages I and II of the collapse (see figure \ref{fig2}).
Evaluating the $u$ integration on the real line, it is straightforward to show that
\begin{equation}
\alpha\left(\omega,\omega'\right) \approx - 4\, r_g \, \left(1 - \frac{r_g}{R_0}\right)^\frac14 \, \sqrt{\frac{|\omega'|}{\omega}} \, \cos\left[\frac{\pi \, (l+1)}{2} - \omega_- r_g\right] e^{i \omega' u_0} e^{\pi \omega' r_g} e^{2 i \omega' r_g \log{(\omega r_g)}} \Gamma(- 2 i \omega' r_g).
\end{equation}
Using the above results, the expressions for the $J_{v}^{(l)}$ and $J_{u}^{(l)}$ are
\bqa
J_{v}^{(l)} \approx \left(1 - \frac{r_g}{R_0}\right)^\frac12 \, \int_m^{\infty} \frac{d\omega}{2\pi}\, \omega
\eqa
and
\bqa\label{Jull}
J_{u}^{(l)} \approx \int_m^{\infty} \frac{d\omega}{2\pi} \, \int_{|\omega'| > m} \frac{d\omega'}{2\pi} \, \int_{|\omega''|>m} \frac{d\omega''}{2\pi} \, \frac{\omega' \, \omega''}{\sqrt{|\omega' \, \omega''|}} \, \alpha(\omega,\omega')\, \alpha^*(\omega, \omega'') \, e^{- i \, (\omega' - \omega'')\, u}.
\eqa
The integral over $\omega$ in the expression for $J_{u}^{(l)}$ can first be evaluated as follows:
\bqa\label{alphaalpha}
\int_m^\infty \frac{d\omega}{2\pi} \, \alpha(\omega,\omega') \, \alpha^*(\omega,\omega'') &\approx 8 \, r_g^2 \, \left(1 - \frac{r_g}{R_0}\right)^\frac12 \, \sqrt{|\omega' \omega''|} e^{i (\omega'- \omega'')u_0} e^{\pi (\omega' + \omega'') r_g} \Gamma(- 2i \omega' r_g) \Gamma(2 i \omega'' r_g) \notag \\[2mm]   & \hspace{40mm} \int_{\log{(mr_g)}}^{\infty} \frac{d \left(\log{(\omega r_g)}\right)}{2\pi} e^{2i (\omega' - \omega'') r_g \log{(\omega r_g)}},
\eqa
where, since $\omega_- r_g\gg 1$, we have replaced $\cos^2\left[\frac{\pi \, (l+1)}{2} - \omega_- r_g\right]$ by $1/2$ in the integral---other contributions lead to terms that decay as powers of $1/u$ in equation (\ref{Jull}). The integral above is over the real half-line, hence
\begin{align}
\int_m^\infty \frac{d\omega}{2\pi} \, \alpha(\omega,\omega') \, \alpha^*(\omega,\omega'') &\approx 2 \, r_g \, \left(1 - \frac{r_g}{R_0}\right)^\frac12 \, |\omega'| \, e^{2 \, \pi \omega' r_g} \left| \Gamma( 2i \omega' r_g) \right|^2 \delta{(\omega' - \omega'') }
+ {\rm regular\; term} , \notag \\
&= 2\, \pi \, \left(1 - \frac{r_g}{R_0}\right)^\frac12 \, n(- \omega') \, \delta\left(\omega' - \omega''\right) + {\rm regular \,\, term}, \label{arelat}
\end{align}
where
\begin{equation} \label{ndef}
 n(\omega) = \frac{sign(\omega)}{e^{4\pi r_g \omega} - 1}.
\end{equation}
The ``regular term'' on the right handside of equation \eqref{arelat} is of the form $\textup{p.v.} \left(\frac{i}{\omega'-\omega''}\right)$, hence for large $u$ its contribution to $J_u^{(l)}$, \eqref{Jull}, is negligible.

Substituting equation (\ref{arelat}) into $J^{(l)}_u$ and using $n(-\omega) =  n(\omega) + sign(\omega)$, we obtain the following expression for the flux:
\bqa
J_{u}^{(l)} \approx 2 \left(1 - \frac{r_g}{R_0}\right)^\frac12 \, \left(\int_m^{\infty} \frac{d\omega}{2\pi}\, \omega \, n(\omega) +  \int_m^{\infty} \frac{d\omega}{2\pi}\, \frac{\omega}{2} \right).
\eqa
Therefore, the total flux is
\bqa
J\left(r\approx r_g , t\right) = \sum_l (2l+1)\, \left[J_{u}^{(l)} - J_{v}^{(l)}\right] \approx \left(1 - \frac{r_g}{R_0}\right)^\frac12 \, \sum_l (2\, l + 1)\, \int_m^{\infty} \frac{d\omega}{2\pi} \omega \, n(\omega).
\eqa
This is the black body radiation appearing in the vicinity of the collapsing shell. The flux at infinity is modified due to the mass of the field under consideration and is multiplied by the grey body factor, $\left|T_{\omega,l}\right|^2$, due to the potential barrier (\ref{poten}),
\bqa\label{fluxfinal}
J\left(r\to \infty , t\right) \approx 2  \left(1 - \frac{r_g}{R_0}\right)^\frac12 \, \sum_l (2\, l + 1)\, \int_m^{\infty} \frac{d\omega}{2\pi} \frac{\omega}{e^{4 \pi r_g \omega}-1} \, \left|T_{\omega, l}\right|^2,
\eqa
where at large $l$ the grey body factor $T_{\omega, l}$ can be estimated to behave as $\frac{\left(i \, r_g \, \omega\right)^{l+1}}{\left(2\, l - 1\right)!!}$ (see, for example, reference \cite{Misner:1974qy}).

If we naively set $m=0$ in
equations (\ref{expanh})--(\ref{arelat}), we encounter divergent integrals at $\omega = 0$.
However, in the massless case one has to use
\begin{align} \label{hIIIm0}
\bar{h}_{\omega,l}(r,t) &\approx \sqrt{\pi r_g} \, J_{l+\frac{1}{2}}(\omega_- r_g) \, e^{-\frac{u - u_0}{2r_g}}  + \, \frac{A_{\omega}}{r_g} \,  e^{-i \, \omega \, v }
\end{align}
instead of equation (\ref{hIII}). It is not hard to show that because of the peculiar behavior of the Bessel functions at small values of their argument, $J_{l+ \frac{1}{2}}(x) \sim x^{l + \frac{1}{2}}$, all $\omega$ integrals in equations (\ref{expanh})--(\ref{fluxfinal}) are convergent at their lower bounds. Then the calculation of the flux continues in the same way as for the massive case and we obtain equation (\ref{fluxfinal}) where $m$ is set to zero.

%
%
%
%

\section{Loop corrections and the secular growth} \label{sec:loop}

In the previous section we used the two point Wightman function to calculate the expectation value of the stress--energy tensor. As motivated in the introduction section, here we show that loop corrections to this Wightman function grow with time.

The interaction term for $\lambda \phi^4$ theory in the collapsing shell background has the following form
\bqa
V(t) = \frac{\lambda}{4!} \int \limits^\infty_{R(t)} \phi^4 r^2 dr d\Omega + \frac{\lambda}{4!}\left(\frac{\partial t_-}{\partial t} \right) \int \limits^{R(t)}_0 \phi^4 r^2 dr d\Omega.
\eqa
In the limit $t \to \infty$, the second term is exponentially suppressed because $\frac{\partial t_-}{\partial t} \propto e^{-t/r_g}$. Hence, we can neglect this term. Furthermore, in this limit, we can neglect the difference between $R(t)$ and $r_g$ as the integrals are regular at $r=r_g$.

As mentioned in the section \ref{intro}, the one-loop, bubble diagram correction to the Keldysh propagator does not introduce a secularly growing contribution, hence we consider the next loop order. The two-loop sunset diagram contribution to the Keldysh propagator can be expressed as follows:
\begin{align} \label{DK02}
D^K_{0+2}(1,2) &= \sum \limits_{l_1,m_1,l_2,m_2} Y_{l_1,m_1}(\Omega_1) Y_{l_2,m_2}(\Omega_2) \int\frac{d\omega_1}{2\pi} \frac{d\omega_2}{2\pi} \notag \\[1mm]
& \hspace{20mm} \Bigg\{\left[N_{\omega_1, l_1, n_1|\omega_2, l_2, n_2}(t)+ \delta_{l_1 l_2} \, \delta_{m_1 m_2} \, \delta(\omega_1 - \omega_2 )\right] \bar{h}^*_{\omega_1,l_1}(t_1, r_1) \, \bar{h}_{\omega_2,l_2}(t_2,r_2) \notag \\
& \hspace{45mm} + K_{\omega_1, l_1, n_1|\omega_2, l_2, n_2}(t) \, \bar{h}_{\omega_1,l_1}(t_1,r_1) \, \bar{h}_{\omega_2,l_2}(t_2,r_2) + {\rm h.c.} \Bigg\},
\end{align}
where we have used the definition of the Keldysh propagator in terms of $D_{+-}$ and $D_{-+}$ propagators, equation \eqref{Keldprop}. Also, from equation \eqref{Dpmprop},
\begin{align}
 D_{+-}(1,2)= \sum\limits_{l,m} \int_{m}^{\infty} \frac{d \omega}{2 \pi} \, \bar{h}^*_{\omega,l}(t_1, r_1) \bar{h}_{\omega,l}(t_2, r_2) Y_{l,n}\left(\theta_1,\varphi_1\right) Y_{l,n}\left(\theta_2,\varphi_2 \right), \\
  D_{-+}(1,2)= \sum\limits_{l,m} \int_{m}^{\infty} \frac{d \omega}{2 \pi} \, \bar{h}_{\omega,l}(t_1, r_1) \bar{h}^*_{\omega,l}(t_2, r_2) Y_{l,n}\left(\theta_1,\varphi_1\right) Y_{l,n}\left(\theta_2,\varphi_2 \right).
\end{align}
Neglecting the difference between $t_1$ and $t_2$ and letting $t = (t_1 + t_2)/2$, it can be shown that the coefficients $N$ and $K$ in the Keldysh propagator take a simple form,
\begin{align} \label{nkbh}
N_{\omega,l,n|\omega',l',n'}(t) &= \frac{\lambda^2}{3}\iint^{t}_{t_0} dt_3 \,dt_4 \iint^\infty_{r_g} (r_3 r_4)^2 \, dr_3 \, dr_4  \, \bar{h}_{\omega,l}(r_3,t_3) \, \bar{h}^*_{\omega',l'}(r_4,t_4)\notag \\[2mm]
& \hspace{45mm}   Y(l,n,l',n') \, \prod\limits_{j=1}^3 \int^\infty_m \frac{d\omega_j}{2\pi} \, \bar{h}_{\omega_j,l_j}(r_3,t_3) \, \bar{h}^*_{\omega_j,l_j}(r_4,t_4)
\end{align}
and
\begin{align}\label{nkbh1}
K_{\omega,l,n|\omega',l',n'}(t) &= -\frac{\lambda^2}{3}\int\limits^{t}_{t_0} dt_3 \int\limits^{t_3}_{t_0} dt_4 \iint^\infty_{r_g} (r_3 r_4)^2 \, dr_3 \, dr_4  \, \left\{\bar{h}^*_{\omega,l}(r_3,t_3)\,\bar{h}^*_{\omega',l'}(r_4,t_4)+\left(\omega,l \leftrightarrow \omega',l'\right)\right\} \notag \\[2mm]
& \hspace{45mm} Y(l,n,l',n') \,\prod\limits_{j=1}^3 \int^\infty_m \frac{d\omega_j}{2\pi} \, \bar{h}^*_{\omega_j,l_j}(r_3,t_3) \, \bar{h}_{\omega_j,l_j}(r_4,t_4),
\end{align}
where
\begin{gather*}
  Y(l,n,l',n')= \sum\limits_{l_{1,2,3}; n_{1,2,3}}\braket{l,n,l_1,n_1,l_2,n_2,l_3,n_3}\braket{l',n',l_1,n_1,l_2,n_2,l_3,n_3},\\[2mm]
  \braket{l,n,l_1,n_1,l_2,n_2,l_3,n_3} = \int d\Omega \, Y_{l,n}(\theta,\varphi) \, Y_{l_1,n_1}(\theta,\varphi) \, Y_{l_2,n_2}(\theta,\varphi) \, Y_{l_3,n_3}(\theta,\varphi).
\end{gather*}
Note that we use the real basis of spherical harmonics.

It is worth noting that $N$ and $K$ from (\ref{nkbh}) themselves do not
receive UV divergent contributions---this can be seen from simple power counting even in flat space-time. However, the Keldysh propagator has the standard UV divergence due to integrations over $\omega$. Our main interest, here, is the contribution from IR modes to $N$ and $K$. Hence, we assume some suitable UV renormalization and assume that all coupling constants acquire their physical (renormalized) values.

We start with the consideration of $N_{\omega,l,n|\omega',l',n'}(t)$. Our goal is to single out the largest contribution to $N$ and $K$ in the infinite future limit. This limit corresponds to $\omega \, e^{-t/r_g} \sim \omega' \, e^{-t/r_g} \to 0$ for fixed $\omega$ and $\omega'$---we will keep only the leading terms in this limit. Note that there are no large contributions to $N$ and $K$ during the stationary stage before collapse because of energy conservation---this was explained in the introduction section. Hence, we can put $t_0$ before $t=0$---before the onset of collapse. Furthermore, as we argued at the end of section \ref{sec:har}, very far from the collapsing shell the behavior of the harmonic functions practically does not change after the start of the collapse. Hence, there are no growing contributions from the region of large $r$. Therefore, the harmonic functions in equation (\ref{nkbh}) can be approximated by (\ref{hIII}), i.e. $\bar{h}_{\omega, l} \approx \bar{h}_{\omega, l}(u) + \bar{h}_{\omega, l}(v)$, where $\bar{h}_{\omega,l}(v)$ is the $v$-dependent part of $\bar{h}_{\omega, l} (r,t)$ given in equation (\ref{hIII}), while $\bar{h}_{\omega,l}(u)$ is its $u$-dependent part. Note that we still keep $\infty$ as the upper limits of integration over $r_{3,4}$, because the integrals are rapidly converging---in fact the integrals converge much faster than if the modes at spatial infinity were used. Moreover, the largest contribution to $N_{\omega,l,n|\omega',l',n'}(t)$ comes from the region of integration over $t_3$ and $t_4$ where $t_3 \gg r_g \, \log\left(r_g \, \omega\right)$ and $t_4 \gg r_g \, \log\left(r_g\,\omega'\right)$. In this region we can neglect the dependence of $\bar{h}_{\omega, l}(u_3)$ and $\bar{h}^*_{\omega', l'}(u_4)$ on $u_3$ and $u_4$, respectively. The $v$-dependent parts of $\bar{h}_{\omega,l}(r_3,t_3)$ and $\bar{h}^*_{\omega',l'}(r_4,t_4)$ lead to subleading contributions.

Using the above simplifications, we find that
\begin{align} \label{nbh}
N_{\omega,l,n|\omega',l',n'}(t) &\approx \frac{2 \, \lambda^2}{3 \, r_g^2 \, \sqrt{\omega \, \omega'}} \,\left(1 - \frac{r_g}{R_0}\right)^\frac12 \, \cos\left[\frac{\pi \, (l+1)}{2} - \omega_- r_g\right] \, \cos\left[\frac{\pi \, (l'+1)}{2} - \omega'_- r_g\right]  \nonumber \\[2mm]
& \qquad \qquad Y(l,n,l',n') \, \int^{t}_{r_g \, \log\left(r_g\,\omega\right)} dt_3 \int^{t}_{r_g \, \log\left(r_g\,\omega'\right)} dt_4 \int^\infty_{r_g} r_3^2 dr_3\int^\infty_{r_g} r_4^2 dr_4  \nonumber \\[3mm]
& \hspace{60mm} \prod\limits_{j=1}^3 \int^\infty_m \frac{d\omega_j}{2\pi} \, \bar{h}_{\omega_j,l_j}(r_3,t_3) \, \bar{h}^*_{\omega_j,l_j}(r_4,t_4).
\end{align}
We can again express the harmonics as sums of $u$ and $v$-dependent parts, hence
\begin{align}
 \int^\infty_m \frac{d\omega_j}{2\pi} \bar{h}_{\omega_j,l_j}(r_3,t_3) \, \bar{h}^*_{\omega_j,l_j}(r_4,t_4) &\approx \int^\infty_m \frac{d\omega_j}{2\pi} \left[\bar{h}_{\omega_j,l_j}(u_3) \, \bar{h}^*_{\omega_j,l_j}(u_4) + \bar{h}_{\omega_j,l_j}(v_3) \, \bar{h}^*_{\omega_j,l_j}(v_4) \right.\notag \\[2mm]
 & \hspace{22mm} \left. + \bar{h}_{\omega_j, l_j}(u_3) \bar{h}^*_{\omega_j, l_j}(v_4) + \bar{h}_{\omega_j, l_j}(v_3) \bar{h}^*_{\omega_j, l_j}(u_4)  \right].
\end{align}
Since, $\bar{h}_{\omega, l}(u)$ is proportional to a rapidly oscillating cosine function, the last two terms in the above expression are negligible.
The first contribution, can again be expanded in Fourier modes, as in the previous section. Using equations \eqref{expanh} and \eqref{arelat},
\bqa
\int^\infty_{m} \frac{d\omega_j}{2\pi}  \bar{h}_{\omega_j,l_j}(r_3,t_3) \bar{h}^*_{\omega_j,l_j}(r_4,t_4) \approx  \left(1 - \frac{r_g}{R_0}\right)^\frac12 \frac{1}{r_g^2}\int_{\omega_j>m} \frac{d\omega_j}{4\pi \omega_j} \left\{ \left[n(-\omega_j) \, e^{- i \omega_j (u_3 - u_4)} \right. \right.\notag \\[3mm]
\left.\left. + n(\omega_j) \, e^{ i \omega_j (u_3 - u_4)}\right] +  \, e^{- i \omega_j (v_3 - v_4)} \right\}  + {\rm subleading \,\, terms}\label{hhint},
\eqa
where $n(\omega)$ is defined in equation \eqref{ndef}.

We can now substitute equation \eqref{hhint} into the expression for $ N_{\omega, l, n|\omega', l', n'}(t)$, (\ref{nbh}). Changing the integration variables from $t_3$ and $t_4$ to $T = (t_3 + t_4)/2$ and $\tau = t_3 - t_4$, we obtain
\begin{align}
N_{\omega, l, n|\omega', l', n'}(t) &\approx \frac{\lambda^2}{12 \, r_g^8 \, \sqrt{\omega \, \omega'}} \,\left(1 - \frac{r_g}{R_0}\right)^2 \, \cos\left[\frac{\pi \, (l+1)}{2} - \omega_- r_g\right] \, \cos\left[\frac{\pi \, (l'+1)}{2} - \omega'_- r_g\right] \nonumber \\[2mm]
& \quad \;\; Y(l,n,l',n') \int^{t}_{0} dT \, \int^{\infty}_{-\infty} d \tau \int^\infty_{r_g} r_3^2 dr_3\int^\infty_{r_g} r_4^2 dr_4  \nonumber \\
& \qquad \prod\limits_{j=1}^3 \int_{\omega_j>m} \frac{d\omega_j}{4\pi \omega_j} \left\{\left[ n(-\omega_j) \, e^{- i \omega_j (\tau - \Delta r)}+ n(\omega_j) \, e^{ i \omega_j (\tau - \Delta r)}\right] +  \, e^{- i \omega_j (\tau + \Delta r)} \right\},
\end{align}
where $\Delta r = r_3 - r_4$. Note that the lower limit of integration over $dT$ is irrelevant for the leading term that grows with time. We have also extended the limits of integration over $\tau$ to $\pm\infty$ because the rapid oscillations of the integrand makes their exact position irrelevant as $t \to \infty$.

Now we can see that the integrand of the $dT$ integral does not depend on $T$ itself.
Hence, the leading contribution is $N \sim \lambda^2 \, t$. As explained in the introduction, it is also possible to see how in a stationary situation this contribution would vanish. In such a case, all the exponentials would have the same sign in their exponent, hence the $\tau$ integration would give $\delta(\omega_1 + \omega_2 + \omega_3)$, which is zero given that $\omega_j$ are positive.

Let us now show that the same growth appears in $K$. Making the same approximations in equation (\ref{nkbh1}) as is made
for the calculation of $N$, we arrive at the following expression:
\begin{gather*}
K_{\omega,l,n|\omega',l',n'} \approx -\frac{4 \lambda^2}{3\sqrt{\omega\omega'} r_g^2}\left(1 - \frac{r_g}{R_0}\right)^\frac{1}{2}\cos\left[\frac{\pi(l+1)}{2} - \omega_- r_g\right]\cos\left[\frac{\pi(l'+1)}{2} - \omega'_- r_g\right]\, Y(l,n,l',n')\\[3mm]
\qquad \qquad \int\limits^t_{r_g \log(\omega r_g)} dt_3\int^{t_3}_{r_g \log(\omega' r_g)} dt_4 \, \int\limits^\infty_{r_g} r_3^2 \, dr_3 \int\limits^\infty_{r_g} r_4^2 \, dr_4
\prod\limits^{3}_{i=1} \int\limits^\infty_m \frac{d\omega_i}{2\pi} \bar{h}^*_{\omega_j,l_j}(r_3,t_3)\bar{h}_{\omega_j,l_j}(r_4,t_4)
\end{gather*}
Then, using equation (\ref{hhint}) and performing the same change of integration variables from $t_3, t_4$ to $T$ and $\tau$ as above, we get that
\begin{align*}
K_{\omega,l,n|\omega',l',n'} &\approx -\frac{\lambda^2}{6\sqrt{\omega\omega'} r_g^8}\left(1 - \frac{r_g}{R_0}\right)^2\cos\left[\frac{\pi(l+1)}{2} - \omega_- r_g\right]\cos\left[\frac{\pi(l'+1)}{2} - \omega'_- r_g\right]\\[2mm]
& \quad Y(l,n,l',n')\int^{t}_0 dT \int^\infty_0 d\tau\int\limits^\infty_{r_g} r_3^2 dr_3 \int\limits^\infty_{r_g} r_4^2 dr_4\\
& \quad \prod \limits^{3}_{j=1} \int_{\omega_j> m} \frac{d\omega_j}{4\pi\omega_j}\left\{\left[n(-\omega_j) e^{-i \omega_j(\tau- \Delta r)}+n(\omega_j) e^{i \omega_j(\tau- \Delta r)}\right]+e^{-i \omega_j (\tau+ \Delta r)}\right\}.
\end{align*}
We again see that the integrand of the integral over $T$ does not depend on $T$ itself. Hence, here the leading contribution is also $K \sim \lambda^2 \, t$.

For the massless case the situation is similar, but instead of using the harmonic of the form given in equation \eqref{hIII}, we must use $\bar{h}$ given in equation \eqref{hIIIm0} to evaluate
\bqa
\int_{0}^{\infty} \frac{d\omega}{2\pi} \bar{h}^*_{\omega,l}(r_3,t_3) \, \bar{h}_{\omega,l}(r_4,t_4).
\eqa
However, as before $J_{l+\frac{1}{2}}(\omega)$ is negligible in the small $\omega$ region, so the only difference with the massive case is that the $\omega$ integrals run from $0$ to $\infty$. Therefore, we obtain the same type of behavior of $N$ and $K$ for massless fields.

\subsection{Loop corrections to the Hawking radiation}

Above we have shown that loop corrections to the two--point functions are not suppressed in comparison with the tree--level contribution. In fact, even if $\lambda^2 \ll 1$ after a long enough time of collapse $\lambda^2 \, t \sim 1$. It means that perturbation theory breaks down. In particular, higher loops bring higher powers of $\lambda^2 \, t$ to $N$ and $K$. Thus, one has to do the resummation of all leading loop corrections. This kind of resummation was done for the case of a strong electric field and massive scalar field theory on de Sitter space-time in \cite{Akhmedov:2014doa}, \cite{Akhmedov:2014hfa}, \cite{Akhmedov:2011pj}, \cite{Akhmedov:2012pa}, \cite{Akhmedov:2012dn}, \cite{Akhmedov:2013vka}, \cite{Akhmedov:2013xka}, \cite{Serreau:2013psa}, \cite{Gautier:2013aoa}. For the present case the resummation will be done elsewhere.

Of course it is too early to draw any conclusions before this resummation has been done.
However, based on the intuition gained from previous examples, we make a few remarks on the physical consequence of these growing corrections. Their presence means that after resummation one will get some specific time--dependence $N_{\omega, l, n|\omega', l', n'}(t)$ and $K_{\omega, l, n|\omega', l', n'}(t)$, which does not necessarily have to be linear. Furthermore, on general physical grounds we expect the presence of a stationary final state for the case under discussion. However, we expect that due to the loop nature of the corrections in question this stationary state will be reached at much latter times than the stationarity of the Hawking tree--level flux is reached.

In any case the presence of $N$ and $K$ means the modification of the Hawking flux. In fact, in the case when $K = 0$ and $N \neq 0$ the equation for the flux is as follows:
\begin{align} \label{Jull1}
 J_{u} &\approx \sum_{l,n} \iint_m^{\infty} \frac{d\omega_1 \, d\omega_2}{(2\pi)^2} \, \int_{|\omega'| > m} \frac{d\omega'}{2\pi} \, \int_{|\omega''|>m} \frac{d\omega''}{2\pi} \sqrt{|\omega' \, \omega''|} \nonumber \\[2mm]
 & \hspace{60mm}\alpha(\omega_1,\omega')\, \alpha^*(\omega_2, \omega'') \, \left\langle\left\{a^\dagger_{\omega_1,l,n}, \, a_{\omega_2,l,n}\right\}\right\rangle \, e^{- i \, (\omega' - \omega'')\, u}.
\end{align}
It is not hard to also restore the contribution of $K$. Now because of the loop corrections we have that
\begin{gather}
\left\langle\left\{a^\dagger_{\omega_1,l,n}, \, a_{\omega_2,l,n}\right\}\right\rangle = \delta(\omega_1 - \omega_2) + 2\, N_{\omega_1,l,n|\omega_2,l,n}(t).
\end{gather}
The above tree--level contribution is obtained when $N=0$.
Note that due to the orthogonality of the spherical harmonics only diagonal in $l$ and $n$ components of $N_{\omega_1,l,n|\omega_2,l',n'}$ will contribute to the flux. Thus, one can see that the presence of $N$ does lead to a modification of the Hawking flux.

\section{Conclusions}

In this paper, we show that, in the case of $\lambda \phi^4$ theory, loop corrections to the Hawking flux are not suppressed given a sufficiently long time---in other words, perturbation theory breaks down. In order to make a definitive conclusion about the fate of Hawking radiation and quantum black holes it is necessary to consider the resummation of the leading corrections from all loops, as done for the case of a strong electric field and massive scalar field theory on de Sitter space-time \cite{Akhmedov:2014doa}, \cite{Akhmedov:2014hfa}, \cite{Akhmedov:2011pj}, \cite{Akhmedov:2012pa}, \cite{Akhmedov:2012dn}, \cite{Akhmedov:2013xka} (see \cite{Akhmedov:2013vka} for a review).  We leave the resummation of the corrections, for the present case, for the future.

The presence of unsuppressed loop corrections leads to a modification of the Hawking flux.
Furthermore, while the tree-level flux $\langle T^r_t \rangle$ is universal, loop corrections to it, i.e. $N$ and $K$, are not universal. Their final value after resummation and as $t \to \infty$ will, in general, depend on initial conditions. At the same time, in some approximation they completely characterize the state of the quantum field theory in the vicinity of the black hole and affect the spectrum of the radiation.
In any case, in this paper we have demonstrated that it is inappropriate and premature to draw any conclusions about black hole evaporation and puzzles associated with it, such as the information paradox \cite{Hawking:1976ra}, based on a study of the tree-level contributions to the energy-momentum tensor.

\vspace{5mm}

\noindent
{\bf{Acknowledgements}} We would like to acknowledge discussions with Mahdi Godazgar, Emil Mottola, Andreas Wipf and Oleg Kancheli. We would like to thank the AEI, in particular Hermann Nicolai and Stefan Theisen, for their generous hospitality while this project was being done.
The work of ETA and FKP was partially supported by the grant for the support of the leading scientific schools SSch--1500.2014.2, by their grants from the Dynasty foundation and by financial support from the Government of the Russian Federation within the framework of the implementation of the 5-100 Programme Roadmap of the National Research University Higher School of Economics. The work of ETA is done under the partial support of the RFBR grant 14-01-90405 Ukr--a. H.G.\ is supported by King's College, Cambridge and acknowledges funding
from the European Research Council under the European Community's
Seventh Framework Programme (FP7/2007-2013) / ERC grant agreement no.
[247252].
The work of FKP is done under the partial support of the RFBR grant 14-02-31446-mol-a and by the support from the Ministry of Education and Science of the Russian Federation (Contract No. 02.A03.21.0003 dated of August 28, 2013) .

\appendix

\section{Normalization of the harmonic functions} \label{app:norm}

In this appendix, we determine the normalization of the harmonics inside the shell before collapse using the condition from the canonical commutation relation. As we remark in section \ref{sec:har}, it is not necessary to fix the normalization in order to find our results---we simply fix the normalization for convenience.
For the modes inside the shell, given by equation \eqref{modeis}, the left handside of condition \eqref{cancomcon} reduces to
\begin{align*}
    & \frac{2 i}{\sqrt{r r'}} \left( 1- \frac{r_g}{R_0} \right)^{-1}
\int_{\omega > m} \frac{d\omega}{2 \pi} \omega \left|{\cal A}_{\omega}\right|^2
J_{l+\frac{1}{2}} \left(\sqrt{ \omega_-^2 -m^2} \; r \right)
J_{l+\frac{1}{2}} \left(\sqrt{ \omega_-^2 - m^2} \; r' \right) \\[3mm]
   & \hspace{60mm} = \frac{i}{\pi \sqrt{r r'}} \int_{p > m \left( 1- \frac{r_g}{R_0}
\right)^{-1/2} \sqrt{\frac{r_g}{R_0}}} d p p \left|{\cal A}_{p}\right|^2
J_{l+\frac{1}{2}} \left(p r \right) J_{l+\frac{1}{2}} \left(p r' \right),
\end{align*}
where we have made use of the following substitution for $\omega$:
$$
p= \sqrt{\omega_-^2 -m^2}.
$$
When $m \, r_g \ll \left( 1- \frac{r_g}{R_0} \right)^{1/2}$, we can use the
normalization identity for Bessel functions whereupon we
obtain~\footnote{We deduce that $A_{\omega(p)}$ must be $p$-independent
in order to obtain the form of the normalization identity for Bessel
functions.}
\bqa
\frac{1}{\pi r^2}  \left|{\cal A}_\omega\right|^2 \delta(r-r').
\eqa
Hence,
\bqa \label{Anorm}
{\cal A}_{\omega} \approx \sqrt{\pi}.
\eqa

\section{The boundary condition for the harmonic functions on the shell before collapse} \label{app:ABcoeff}

We fix the coefficients $A_\omega$ and $B_\omega$ of the harmonic modes in the vicinity of the shell, \eqref{modeos}, by imposing the boundary condition \eqref{bound} on the shell.  First, consider the continuity of the modes at the shell, \eqref{statbound}. From $|R_0 - r_g| \ll r_g,$ we deduce that
\bqa \left| 1- \frac{r_g}{R_0} \right|^{-1} \gg \frac{R_0}{r_g} \approx 1.
\eqa
Therefore, from equation \eqref{modeis},
\begin{align*}
h_{\omega, l}\left[R_0^{\phantom{\frac12}} - \,\, 0 \right] &\approx \sqrt{\pi} \frac{1}{\sqrt{R_0}}
J_{l+\frac{1}{2}} \left(\omega \left(1-\frac{r_g}{R_0}\right)^{-1/2} R_0 + \mathcal{O}\left(\sqrt{1-\frac{r_g}{R_0}} \right) \right), \\[3mm]
&\approx \sqrt{\frac{2}{\omega}} \frac{1}{R_0} \left(1-\frac{r_g}{R_0}\right)^{1/4} \sin\left(\omega \left(1-\frac{r_g}{R_0}\right)^{-1/2} R_0 - \frac{\pi l}{2} \right) + \mathcal{O}\left(\left(1-\frac{r_g}{R_0}\right)^{3/4}\right).
\end{align*}
Hence, up to lower order terms in  $\left(1- \frac{r_g}{R_0}\right)$, the continuity of the $\phi_l$ across the shell gives
\bqa \label{alphaccon}
\sqrt{\frac{2}{\omega}} \left(1-\frac{r_g}{R_0}\right)^{1/4} \sin\left(\omega \left(1-\frac{r_g}{R_0}\right)^{-1/2} R_0 - \frac{\pi l}{2} \right) \approx A_\omega e^{-i \omega  R_0^* } + B_{\omega} e^{i \omega R_0^* },
\eqa
which, in particular, implies that $B_\omega = A_\omega^*.$
Furthermore, the second boundary condition in \eqref{statbound} for the radial derivative of $\phi_l$, again up to lower order terms in  $\left(1- \frac{r_g}{R_0}\right)$, gives
\bqa \label{alphadccon}
i \sqrt{\frac{2}{\omega}} \left(1-\frac{r_g}{R_0}\right)^{1/4} \cos\left(\omega \left(1-\frac{r_g}{R_0}\right)^{-1/2} R_0 - \frac{\pi l}{2} \right) \approx  A_\omega e^{-i \omega R_0^* } - A_\omega^* e^{i \omega R_0^* }.
\eqa

Using the above equations, it is straightforward to see that
\bqa
A_\omega = B_{\omega}^* = \frac{i^{l+1}}{ \sqrt{2\, \omega}} \left( 1- \frac{r_g}{R_0} \right)^{1/4}  e^{i \, \omega \, \left[R_0^* - R_0\, \left(1-\frac{r_g}{R_0}\right)^{-\frac12} \right] } + \mathcal{O}\left(\left( 1- \frac{r_g}{R_0} \right)^{3/4} \right).
\eqa

\section{The boundary condition for the harmonic functions on the shell during the late-stage of collapse} \label{app:boundcon}

In this appendix, we show that the solution for the harmonic function outside and in the vicinity the shell, \eqref{f1}, which was found in section \ref{harmiii} by imposing continuity at the shell also satisfies the boundary condition on the normal derivative, equation \eqref{dbconinf}.

In section \ref{harmiii} we added a constant term to $r h_{\omega, l}$ inside the shell because of the approximation $\nu \approx 1$. In the boundary condition for the normal derivatives of the harmonics this term simply cancels the $v$-dependent term in $r h_{\omega, l}$ outside the shell. This makes sense because the boundary condition is a condition on derivatives normal to the shell, namely the $u$-dependence of the harmonics outside the shell. In what follows, we will concentrate on the non-constant piece inside the shell and the $u$-dependent term in the harmonics outside the shell.

Using the harmonic function inside the shell, given in equation \eqref{modeiii}, the left-handside of equation \eqref{dbconinf} is
\begin{align}
&\left(\frac{\partial t_-}{\partial t} \right) \left[\left(\nu \, \partial_{t_-} -  \partial_r \right) h_{\omega, l}\right]_{r=R(t) - 0} \notag \\[2mm]
&= \frac{R_0 - r_g}{\nu r_g}\frac{\sqrt{\pi}}{\sqrt{R}} e^{-\frac{t}{r_g}} e^{- i \omega_- \frac{R_0 - r_g}{\nu}\left(1 - e^{-\frac{t}{r_g}}\right)} \left[\left(\frac{1}{2 R} - i \omega_- \nu \right)   J_{l+\frac{1}{2}} \left(\sqrt{\omega_-^2 -m^2} \; R\right) \right. \notag \\[3mm]
& \hspace{39mm} \left. - \frac{1}{2 } \sqrt{\omega_-^2- m^2} \left( J_{l-\frac{1}{2}} \left(\sqrt{\omega_-^2 -m^2} \; R\right) - J_{l+\frac{3}{2}} \left(\sqrt{\omega_-^2 -m^2} \; R\right) \right) \right], \label{bounconlhs}
\end{align}
where we have used the expression for $t_-$, \eqref{rel2}, as $t \longrightarrow \infty$.
The right-handside of equation \eqref{dbconinf} is
\begin{align}
2 \left[ \partial_{u} h_{\omega, l} \right]_{r=R(t) + 0}
&= - \frac{\sqrt{\pi}}{2 R^{3/2}} R'(u) e^{- i \omega_- \frac{(R_0 -r_g)}{\nu} \left( 1- e^{-\frac{u+ R_0^* +r_g}{2 r_g}}\right)} \left[2 J_{l+\frac{1}{2}} \left(\sqrt{\omega_-^2 -m^2} \; R(u) \right) \right. \notag \\[4mm]
& \hspace{15mm} \left. - 2  R\sqrt{\omega_-^2 -m^2}  \left(  J_{l-\frac{1}{2}} \left(\sqrt{\omega_-^2 -m^2} \; R\right) - J_{l+\frac{3}{2}} \left(\sqrt{\omega_-^2 -m^2} \; R\right) \right) \right]  \notag \\[3mm]
& \quad - 2 i \omega_-   \frac{R_0 -r_g}{2\nu r_g \sqrt{R} } \sqrt{\pi} \, e^{-\frac{t}{r_g}}  J_{l+\frac{1}{2}} \left(\sqrt{\omega_-^2 -m^2} \; R \right) e^{- i \omega_- \frac{(R_0 -r_g)}{\nu} \left( 1- e^{-\frac{u+ R_0^* +r_g}{2 r_g}}\right)}.
\end{align}
where we have used equation \eqref{Ru},
\bqa
R'(u) =  - \frac{R_0-r_g}{2 r_g} e^{-\frac{t}{r_g}},
\eqa
and the behavior of $u$ on the shell as $t \longrightarrow \infty$, \eqref{uvapp}. Using the above relation and equation \eqref{uvapp}, the right-handside of equation \eqref{dbconinf} simplifies to
\begin{align}
2 \left[ \partial_{u} h_{\omega, l} \right]_{r=R(t) + 0}
&= \frac{R_0 - r_g}{r_g}\frac{\sqrt{\pi}}{\sqrt{R}} e^{-\frac{t}{r_g}} e^{- i \omega_- \frac{R_0 - r_g}{\nu}\left(1 - e^{-\frac{t}{r_g}}\right)} \left[\left(\frac{1}{2 R} -  \frac{i \omega_-}{\nu} \right)   J_{l+\frac{1}{2}} \left(\sqrt{\omega_-^2 -m^2} \; R\right) \right. \notag \\[3mm]
&  \hspace{11mm} \left. - \frac{1}{2 } \sqrt{\omega_-^2- m^2} \left( J_{l-\frac{1}{2}} \left(\sqrt{\omega_-^2 -m^2} \; R\right) - J_{l+\frac{3}{2}} \left(\sqrt{\omega_-^2 -m^2} \; R\right) \right) \right].
\end{align}
Comparing the above equation to equation \eqref{bounconlhs}, we conclude that the boundary condition on the normal derivatives is satisfied if $\nu =1.$

\end{document}